\documentclass[12pt]{article}
\pdfoutput=1
\usepackage{putex}

\usepackage[normalem]{ulem}

\makeatletter
\renewcommand*\env@matrix[1][*\c@MaxMatrixCols c]{%
  \hskip -\arraycolsep
  \let\@ifnextchar\new@ifnextchar
  \array{#1}}
\makeatother

\begin{document}

\preprint{PUPT-2479}

\institution{Stanford}{Stanford Institute for Theoretical Physics, Stanford University, Stanford, CA 94305, USA}
\institution{PCTS}{Princeton Center for Theoretical Science, Princeton University, Princeton, NJ 08544, USA}
\institution{PU}{Joseph Henry Laboratories, Princeton University, Princeton, NJ 08544, USA}
\institution{Harvard}{Department of Physics, Harvard University, Cambridge, MA 02138, USA}
\institution{Perimeter}{Perimeter Institute for Theoretical Physics, Waterloo, Ontario N2L 2Y5, Canada}

\title{Scaling dimensions of monopole operators in the $\CP^{N_b - 1}$ theory in $2+1$ dimensions}

\authors{Ethan Dyer,\worksat{\Stanford} M\'ark Mezei,\worksat{\PCTS}  Silviu S.~Pufu,\worksat{\PU} and Subir Sachdev\worksat{\Harvard,\Perimeter}}

\abstract{
We study monopole operators at the conformal critical point of the $\CP^{N_b - 1}$ theory in $2+1$ spacetime dimensions.  Using the state-operator correspondence and a saddle point approximation, we compute the scaling dimensions of these operators to next-to-leading order in $1/N_b$.  We find remarkable agreement between our results and numerical studies of quantum antiferromagnets on two-dimensional lattices with SU($N_b$) global symmetry, using the mapping of the monopole operators to valence bond solid order parameters of the lattice antiferromagnet.
}

\date{}

\maketitle

\tableofcontents

\section{Introduction}

In $2+1$ dimensions, pure U(1) gauge theory confines \cite{Polyakov:1975rs}. One can prevent confinement by introducing a sufficiently large number $N$ of massless matter fields, in which case the infrared dynamics is believed to be governed by a non-trivial interacting conformal field theory (CFT).   Such CFTs arise quite frequently in the description of quantum critical points of condensed matter systems in two spatial dimensions \cite{Wen:1993zza, Chen:1993cd, Sachdev97, Rantner01, Rantner:2002zz, Motrunich:2003fz, SVBSF, SBSVF, Hermele, Hermele05, Ran06, Kaul08, Kaul:2008xw, Sachdev:2010uz}.  They also serve as useful toy-models for more intricate four-dimensional dynamics, as they can be studied perturbatively in the $1/N$ expansion, where the gauge interactions are suppressed \cite{Appelquist:1981vg, Appelquist:1986fd, Appelquist:1988sr, Appelquist:1981sf}.

Our goal in this paper is to study monopole operators in one such CFT, namely the $\CP^{N_b - 1}$ theory tuned to criticality.  This theory is a nonlinear sigma-model with $\CP^{N_b - 1}$ target space, and can be equivalently described as a U(1) gauge theory coupled to $N_b$ complex scalars of unit charge that satisfy a length constraint; see~\cite{coleman1988aspects} for a textbook treatment.  The action is 
 \es{CPNAction}{
   \mathcal{S} = \frac{N_b}{\mathfrak{g}} \int d^3 x  \Bigl[ \abs{(\partial_\mu - i A_\mu) \phi_\alpha}^2  + i \lambda ( \abs{\phi_\alpha}^2 - 1) \Bigr] \,, 
 }
where $\alpha = 1 \ldots N_b$, $\lambda$ is a Lagrange multiplier imposing the length constraint, and $\mathfrak{g}$ is a coupling constant.  This theory becomes critical provided that one tunes the coupling to $\mathfrak{g} = \mathfrak{g}_c$ for some $\mathfrak{g}_c$.  

The interest in monopole operators in this theory is motivated by their interpretation as order parameters for the valence bond solid (VBS) order of quantum antiferromagnets \cite{Read:1990zza,Read:1989zz,Murthy:1989ps}.  The quantum antiferromagnets are defined on bipartite lattices in two spatial dimensions, and have a global 
SU($N_b$) symmetry. Each site of the first (second) sublattice has states transforming under the fundamental (anti-fundamental) of SU($N_b$). The sites interact via 
short-range exchange interactions with SU($N_b$) symmetry. There is no explicit reference to a gauge field in the lattice Hamiltonian. Nevertheless, when the spin states
on each site are represented in terms of `parton' degrees of freedom, a U(1) gauge field, $A_\mu$, emerges in the path integral formulation in the $1/N_b$ expansion.
The partons become the $\phi_\alpha$ matter fields in this gauge theory.
As their exchange constants are varied, such antiferromagnets can exhibit ground states with two distinct broken symmetries. First, there is the state with antiferromagnetic
order, in which the SU($N_b$) symmetry is broken by the condensation of $\phi_\alpha$: this is the Higgs phase of the U(1) gauge theory. Second, we have the 
state with VBS order in which SU($N_b$) symmetry is preserved but a lattice rotation symmetry is broken. In the U(1) gauge theory, this state appears initially as a Coulomb phase;
however, the confinement of the U(1) gauge theory by the proliferation of monopoles leads to the appearance of VBS order in the lattice antiferromagnet. This is a consequence of subtle Berry phases associated with the monopole tunneling events in the lattice antiferromagnet, which endow the monopole operators with non-trivial transformations under lattice symmetry operations, identical to those of the VBS order. The quantum phase transition between these two states of the lattice antiferromagnet has been argued
to be continuous \cite{SVBSF,SBSVF}, and described by the $\CP^{N_b - 1}$ theory in (\ref{CPNAction}), with monopoles suppressed at the quantum critical point;
the critical point is therefore `deconfined'.

From the perspective of the $\CP^{N_b - 1}$ theory, this connection to the VBS order of the antiferromagnet is powerful because it allows the monopole operators
to be expressed as simple, local, gauge-invariant operators of the lattice model. Moreover, the couplings of the lattice model can be chosen
to avoid the `sign' problem of quantum Monte Carlo, and this allows efficient studies on large lattices of the $\CP^{N_b - 1}$ CFT at the deconfined critical point between the Higgs and VBS phases \cite{2009PhRvB..80r0414L, 2012PhRvL.108m7201K, 2013PhRvL.111h7203P, 2013PhRvL.111m7202B, 2015arXiv150205128K}. 
Block {\em et al.\/} \cite{2013PhRvL.111m7202B, 2015arXiv150205128K} have obtained the scaling dimensions of VBS operators on a number of lattice antiferromagnets, 
and here we will compare their results with the $1/N_b$ expansion for the scaling dimensions of the monopole operators in the $\CP^{N_b - 1}$ CFT.

Unlike the lattice antiferromagnet, the monopole operators of the $\CP^{N_b - 1}$ field theory are not defined simply as products of fields that appear in the Lagrangian~\eqref{CPNAction}. Instead, the monopole operators appear as singular boundary conditions that these fields must obey at the point where the monopole operator is inserted \cite{Murthy:1989ps,Borokhov:2002ib, Borokhov:2002cg}.
One way of defining monopole operators is the following.  The $\CP^{N_b -1}$ model has SU($N_b$)$ \times$ U(1)$_\text{top}$ global symmetry.  Under the action of SU($N_b$), the charged scalar fields $\phi_\alpha$ transform in the fundamental representation.  The U(1)$_\text{top}$ factor in the global symmetry group is a topological symmetry whose conserved current is 
 \es{ConsCurrent}{
   j_\mu = \frac{1}{8 \pi} \epsilon_{\mu\nu\rho} F^{\nu \rho} \,.
 }
The Dirac quantization condition implies that the conserved charge $q = \int d^2 x \, j_0$ satisfies $q \in \Z/2$. It should be noted that our definition of $q$ here differs by a factor of 2 from earlier work \cite{Murthy:1989ps, Metlitski:2008dw, 2013PhRvL.111m7202B}. 

One can define the monopole operators as operators that have non-vanishing U(1)$_\text{top}$ charge $q$.  For each $q$, we will focus on the monopole operator ${\cal M}_q$ with the lowest scaling dimension.  The other operators in the same topological charge sector can be thought intuitively as products between the monopole operator with lowest scaling dimension and more conventional operators from the $q=0$ sector.   Throughout this paper, we assume without loss of generality that $q \geq 0$.  The physical quantities that we compute depend only on $\abs{q}$.

We aim to determine the scaling dimension of the monopole operator ${\cal M}_q$ with U(1)$_\text{top}$ charge $q$ to next-to-leading order in $1/N_b$.  The most convenient way of performing this computation is to use the state-operator correspondence, under which a local operator inserted at the origin of $\R^3$ is mapped to a state of the CFT on the conformally-flat background $S^2 \times \R$.  The scaling dimension of the monopole operator ${\cal M}_q$ is therefore mapped to the ground state energy on $S^2$ in the sector with magnetic flux $\int F = 4 \pi q$ through the $S^2$ \cite{Borokhov:2002cg,Borokhov:2002ib,Metlitski:2008dw}. (See also \cite{Murthy:1989ps} for an earlier approach to computing scaling dimensions of monopole operators.)  As is standard in thermodynamics, this ground state energy $\Delta_q$ can be related to the partition function on $S^2 \times \R$ via
 \es{PartFn}{
  \Delta_q = -\lim_{\beta \to \infty} \frac{1}{\beta} \log Z_q(\beta) \equiv -\log Z^{S^2 \times \R}_q \equiv {\cal F}_q \,,
 }
where in the middle equality we regularized the $S^2 \times \R$ partition function by compactifying the $\R$ direction into a large circle of circumference $\beta$.

In the case at hand, the $S^2$ ground state energy in the presence of $4 \pi q$ magnetic flux is easily computed at leading order in $N_b$, where the Lagrange multiplier field $\lambda$ and the gauge field don't fluctuate and assume a saddle point configuration that minimizes this energy.  It is reasonable to assume that the large $N_b$ saddle point (for both the magnetic flux through $S^2$ and the value assumed by the Lagrange multiplier field) is rotationally-symmetric.  At leading order in $N_b$, the ground state energy on $S^2$ comes from performing the Gaussian integral over the matter fields \cite{Murthy:1989ps}.  The $1/N_b$ correction to this result takes into account the Gaussian fluctuations of the gauge field as well as of those of $\lambda$.  We perform an analysis of these fluctuations around the rotationally-invariant saddle point;  by computing their determinant, we extract the $1/N_b$ correction to the scaling dimension of ${\cal M}_q$. Our results are listed in Table~\ref{tab:intro} below.
 \begin{table}[!htp]
\begin{center}
\begin{tabular}{c|c|c}
$q$ & $ \Delta_q$ & $N_b$ for which $\Delta_q < 3$\\
\hline
$0$ & $0$  & $< \infty$ \\
$1/2$ &  $0.1245922\,N_b+0.3815+O(N_b^{-1})$& $\leq 21$ \\
$1$ &  $0.3110952\,N_b+0.8745+O(N_b^{-1})$ & $\leq 6$\\
$3/2$ &  $0.5440693\,N_b+1.4646+O(N_b^{-1})$ & $\leq 2$\\
$2$ &  $0.8157878\,N_b+2.1388+O(N_b^{-1})$ & none \\
$5/2$ &  $1.1214167\,N_b+2.8879+O(N_b^{-1})$ & none \\
\end{tabular}
\end{center}
\caption{Results of the large $N_b$ expansion of the monopole operator dimensions $\Delta_q$ obtained through calculating the ground state energy in the presence of $2q$ units of magnetic flux through $S^2$. In the last column of the table we listed our estimates for when the monopole operators are relevant.  
}
\label{tab:intro}
\end{table}%

\begin{figure}[h]
\begin{center}
\includegraphics[width=5in]{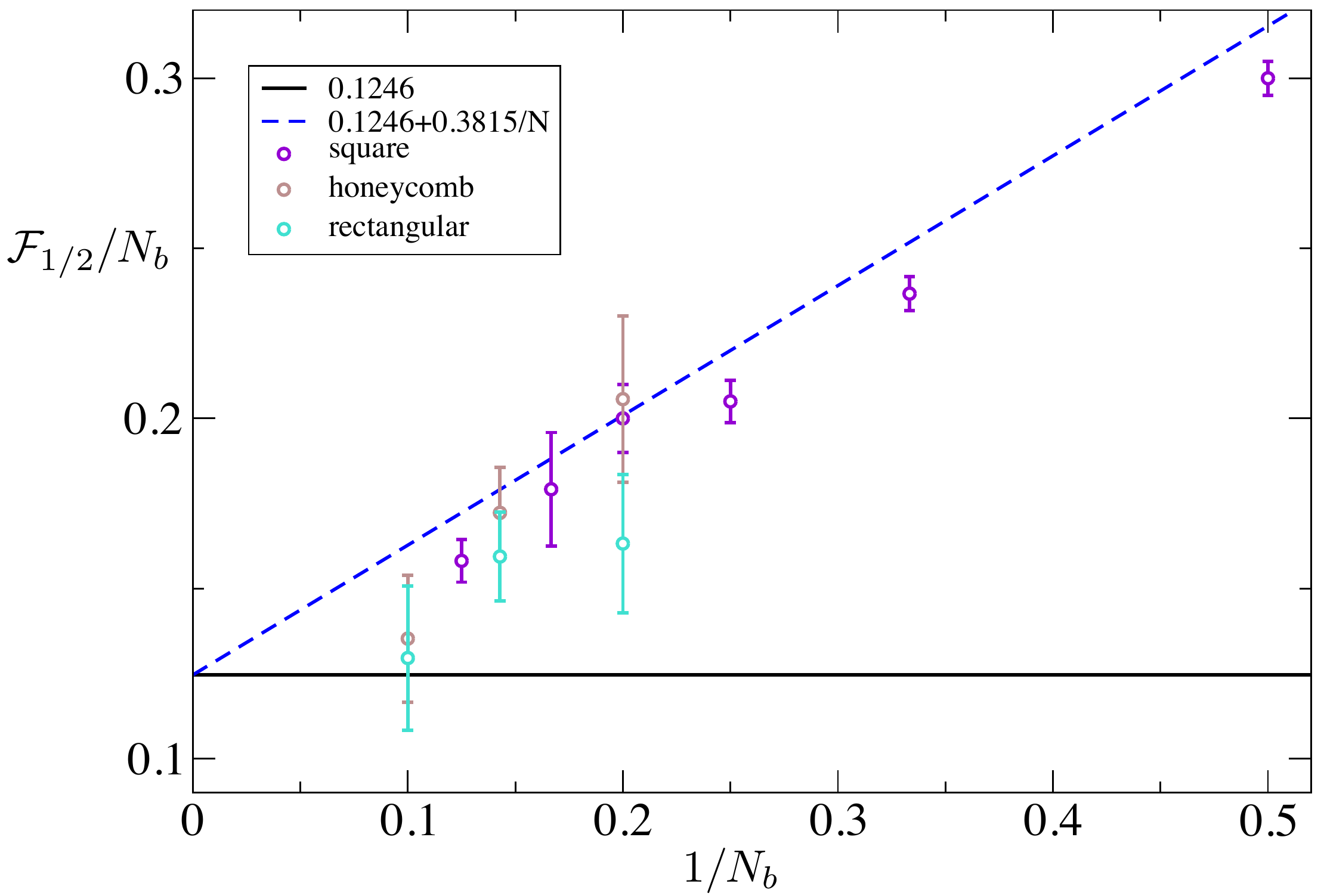}
\end{center}
\caption{The scaling dimension of the $q=1/2$ monopole operator, $\mathcal{F}_{1/2}$. 
The solid line is the $N_b = \infty$ result (Ref.~\cite{Murthy:1989ps}), and the dashed line is the leading $1/N_b$ correction computed in the present paper
(see Table~\ref{tab:intro}).
The quantum Monte Carlo results are for lattice antiferromagnets with global SU($N_b$) symmetry on the square (Refs.~\cite{2009PhRvB..80r0414L, 2012PhRvL.108m7201K}), honeycomb (Ref.~\cite{2013PhRvL.111m7202B}), and 
rectangular (Ref.~\cite{2013PhRvL.111m7202B}) lattices.
}
\label{fig:qmc}
\end{figure}

In Figure~\ref{fig:qmc} we compare our result for $\mathcal{F}_{1/2}$  to numerical studies of the lattice antiferromagnet and find a remarkable agreement.  This agreement suggests that the next correction to $\mathcal{F}_{1/2}$ in $1/N_b$ is probably quite small.   From comparing the scaling dimensions collected in Table~\ref{tab:intro} to $3$, we can also estimate the upper bound on $N_b$ below which the monopole operators are expected to be relevant; these bounds are also presented in Table~\ref{tab:intro}.  There is inherently some uncertainty in these estimates, as they come from extrapolating the large $N_b$ expansion to small values of $N_b$. Nevertheless, our relevance bounds come close to what Ref.~\cite{2013PhRvL.111m7202B} found from numerics, as can be seen from Table I in~\cite{2013PhRvL.111m7202B}.

From Table~\ref{tab:intro}, it can be noticed that $ \Delta_q$ grows with $q$ approximately as $q^{3/2}$. In Section~\ref{STABILITY} we give an argument for this behavior, and find that
\es{LargeqIntro}{
 \Delta_q
&= q^{3/2} \, \le(0.26408 \,N_b+ c + O(N_b^{-1}) \ri)+O(q^{1/2})\
}
with a constant $c \approx 0.23$ that can be deduced from the numerical values presented in Table~\ref{tab:intro}.  In Section~\ref{STABILITY}, we take the first steps towards deriving the value of $c$ analytically.  It would be very interesting to develop a more complete understanding of the large $q$ behavior.

Similar calculations were performed in theories with fermionic matter in~\cite{Pufu:2013vpa,Dyer:2013fja}.\footnote{The calculation in the case with scalar matter performed in this paper is technically more challenging than that for fermionic matter due to additional UV divergences.  One can renormalize these divergences using zeta-function techniques, as we do here, but the same result can be derived in other renormalization schemes~\cite{ScalarQED}.} (See also \cite{Pufu:2013eda} where only the fluctuations of the Lagrange multiplier field were calculated with the purpose of studying monopole insertions in theories with global U(1) symmetry.)  In \cite{Sachdev:2012tj, Iqbal:2014cga}, monopole operators were studied holographically. For studies of monopole operators in supersymmetric theories, see, for instance, \cite{Borokhov:2003yu, Benna:2009xd, Gustavsson:2009pm, Benini:2009qs,Benini:2011cma, Aharony:2015pla}.

The rest of this paper is organized as follows.  In Section~\ref{SETUP}, we set up our computation.  In Section~\ref{LEADING}, we review the leading order analysis at large $N_b$.  In Section~\ref{CORRECTIONS}, we examine the $1/N_b$ corrections around the spherically-symmetric saddle points of the effective action for the gauge field and Lagrange multiplier.  In Section~\ref{STABILITY} we analyze the behavior at large $q$.  We end with concluding remarks in Section~\ref{sec:conc}.  Several technical details of our computation are included in the Appendices.


\section{Setup}\label{SETUP}

In order to study the large $N_b$ limit of the $\CP^{N_b - 1}$ theory, it is convenient to rescale the fields such that the action \eqref{CPNAction}, appropriately generalized to that on an arbitrary conformally-flat space with metric tensor $g_{\mu\nu}$,  takes the form
 \es{CPNActionRescaled}{
   \mathcal{S} = \frac{1}{\mathfrak{g}} \int d^3 x \sqrt{g(x)} \Bigl[ g^{\mu\nu} \left[(\nabla_\mu + i A_\mu) \phi_\alpha^\ast \right] \left[ (\nabla_\nu - i A_\nu) \phi_\alpha \right] +{{\cal R}\ov 8}\,|\phi_\alpha|^2+ i \lambda ( |\phi_\alpha|^2 - N_b) \Bigr] \,,
 }
where ${\cal R}$ is the Ricci scalar.\footnote{We could have absorbed the conformal mass term ${{\cal R}\ov 8}\,|\phi_\alpha|^2$ by shifting $ i \lambda$, but we chose not to. } In this paper we will work on $S^2 \times \R$, which we parameterize by coordinates $x \equiv (\theta, \phi, \tau)$.

The monopole scaling dimension is equal to the ground state energy $\mathcal{F}_q$ on $S^2 \times \R$ in the presence of a magnetic flux $\int F = 4 \pi q$ through the $S^2$. Our main task is to determine the $1/N_b$ expansion of this ground state energy, which we write as
 \es{expand}{
   \mathcal{F}_q = N_b \left( \mathcal{F}_q^{\infty} + \frac{1}{N_b} \delta \mathcal{F}_q  + {\cal O}(1/N_b^2)\right) \,.
 }
When $q=0$, the corresponding ground state energy ${\cal F}_0$ is nothing but the scaling dimension of the unit operator.  We therefore must have $\mathcal{F}_0 = 0$.

It is not hard to see that at large $N_b$, the fluctuations of $\lambda$ and of $A_\mu$ around any saddle point configuration are suppressed.  Indeed, upon integrating out the scalars $\phi_\alpha$ in the action \eqref{CPNActionRescaled}, one obtains an effective action for the gauge field and Lagrange multiplier given by
 \es{EffAction}{
  {\cal S}_\text{eff}[A_\mu, i \lambda] =   N_b \left[ \tr \log \left( -(\nabla_\mu - i A_\mu)^2 + \frac 14 + i \lambda\right)   - \frac{i}{\mathfrak{g}} \int d^3 x \sqrt{g} \lambda \right]  \,.
 }
Let's expand $A_\mu$ and $\lambda$ around a saddle point by writing\footnote{In terms of the quantity $a_q^2$ introduced in \cite{Metlitski:2008dw}, we have $a_q^2 = \mu_q^2 -  q^2$.}
 \es{AlambdaExpansion}{
  A_\mu &= {\cal A}_\mu^q + a_\mu \,, \\
  i \lambda &= \mu_q^2  + i \sigma \,,
 }
where $a_\mu$ and $\sigma$ are fluctuations around the saddle point configuration $A_\mu = {\cal A}_\mu$ and $i \lambda = \mu_q^2$.  As can be easily seen from \eqref{EffAction}, the effective action for these fluctuations is proportional to $N_b$, so their typical size is of order $1/\sqrt{N_b}$ and are therefore suppressed at large $N_b$.  To leading order in $1/N_b$, it is therefore correct to set $a_\mu = \sigma = 0$, provided that the background values ${\cal A}_\mu^q$ and $\mu_q^2$ are such that the saddle point conditions
 \es{SaddlePt}{
  \frac{\delta {\cal S}_\text{eff}[A_\mu, i \lambda]}{\delta A_\mu} \biggr|_{\sigma = a_\mu = 0} =  \frac{\delta {\cal S}_\text{eff}[A_\mu, i \lambda]}{\delta \lambda} \biggr|_{\sigma = a_\mu = 0} = 0
 }
are obeyed.  One can then develop the $1/N_b$ expansion to higher orders by integrating over the fluctuations $a_\mu$ and $\sigma$ using the effective action \eqref{EffAction}. 

In this paper we will focus only on saddles that are rotationally-invariant on $S^2$ and translationally-invariant along $\R$.  These conditions imply that $\mu_q^2$ is a constant and that, in the sector of monopole flux $\int F = 4 \pi q$, the background magnetic field ${\cal F}^q = d {\cal A}^q$ is uniformly distributed over $S^2$:
 \es{UnifDistrib}{
  {\cal F}^q = q \sin \theta d\theta \wedge d\phi \,.
 }
One can choose a gauge where the background gauge potential ${\cal A}^q$ can be written as 
 \es{BackA}{
  {\cal A}^q = q (1 - \cos \theta) d\phi \,.
 }
(This expression is well-defined everywhere away from the South pole at $\theta = \pi$.)  The saddle point condition \eqref{SaddlePt} is satisfied provided that the constant $\mu_q^2$ is chosen such that it minimizes the value of the effective action evaluated when $a_\mu = \sigma = 0$.  In other words, the equation that determines $\mu_q^2$ is
 \es{Saddlemu}{
  \frac{\partial {\cal S}_\text{eff}[{\cal A}_\mu^q, \mu_q^2]}{\partial \mu_q^2} = 0 \,.
 }  
This equation depends non-trivially on $q$, and hence so does $\mu_q^2$.

In the next section, we calculate the coefficient $\mathcal{F}_q^{\infty}$ by simply evaluating ${\cal S}_\text{eff}$ at the saddle point, while in Section~\ref{CORRECTIONS}, we compute the correction $\delta \mathcal{F}_q$ from the functional determinant of the fluctuations around this saddle point.


\section{$N_b = \infty$ theory}\label{LEADING}

At leading order in $N_b$, one can identify 
 \es{LeadingNb}{
   N_b \mathcal{F}_q^{\infty} = {\cal S}_\text{eff}[{\cal A}_\mu^q, \mu_q^2] \,,
 }
evaluated for the value of $\mu_q^2$ that solves \eqref{Saddlemu}, or equivalently at large $N_b$
 \es{SaddleLargeN}{
   \frac{\partial {\cal F}_q^\infty}{\partial \mu_q^2} = 0 \,,
 }
and with the coupling $\mathfrak{g}$ tuned to the critical value $\mathfrak{g} = \mathfrak{g}_c$.  In other words,
 \es{LeadingNbAgain}{
  \mathcal{F}_q^{\infty} = \tr \log \left( -(\nabla_\mu - i {\cal A}^q_\mu)^2  + \mu_q^2 + \frac 14\right)   - \frac{4 \pi}{\mathfrak{g}_c} \, \mu_q^2  \,.
 }
Using the fact that the eigenvalues of the gauge-covariant Laplacian on $S^2$ in the presence of magnetic flux $4 \pi q$ are $j (j + 1) - q^2$ \cite{Wu:1976ge,Wu:1977qk}, we obtain
 \es{omegad}{
\mathcal{F}_q^{\infty} =  \int \frac{d \omega}{2 \pi} \sum_{j=q}^{\infty}
(2 j + 1) \log \left[\omega^2 + (j + 1/2)^2 + \mu_q^2 - q^2 \right]   - \frac{4 \pi}{\mathfrak{g}_c} \, \mu_q^2  \,.
 }
The first term in this expression is divergent and requires regularization.  The second term is also divergent because, as we explain shortly, the inverse critical coupling $1/\mathfrak{g}_c$ diverges linearly, and so the second term in \eqref{omegad} cancels part of the divergence in the first term.  To be explicit, let us deduce an expression for $1/\mathfrak{g}_c$.  The saddle point condition \eqref{SaddleLargeN} at $q=0$ can be written as 
 \es{CouplingSaddle}{
  \frac{4 \pi}{\mathfrak{g}} = \int \frac{d\omega}{2 \pi} \sum_{j=0}^\infty (2j+1) \frac{1}{\omega^2 + (j+1/2)^2 + \mu_0^2} \,.
 }
The theory is critical when the correlators on $S^2 \times \R$ are those obtained by conformally mapping the power-law correlators on $\R^3$.  For a scalar field $\phi_\alpha$, this prescription yields a conformally coupled scalar on $S^2 \times \R$, for which $\mu_0 = 0$.  Hence criticality is achieved when, to leading order in $N_b$, we have $\mathfrak{g} = \mathfrak{g}_c$ and $\mu_0^2 = 0$, with
 \es{CouplingSaddleCrit}{
  \frac{4 \pi}{\mathfrak{g}_c} = \int \frac{d\omega}{2 \pi} \sum_{j=0}^\infty (2j+1) \frac{1}{\omega^2 + (j+1/2)^2} \,.
 }

After substituting \eqref{CouplingSaddleCrit} into \eqref{omegad}, the resulting expression is still divergent, but can be rendered finite using, for instance, zeta-function regularization as in \cite{Pufu:2013eda}, or Pauli-Villars regularization as in \cite{Murthy:1989ps, Metlitski:2008dw}.  We will not repeat that calculation here.  The regularized ground state energy coefficient ${\cal F}^\infty_q$ is
 \es{FInfFinal}{
\mathcal{F}_q^{\infty}
 &= 2 \sum_{j=q}^{\infty} \left[ (j + 1/2)  
\left[(j + 1/2)^2 + \mu_q^2 - q^2 \right]^{1/2} - (j+1/2)^2 - \frac{1}{2} (\mu_q^2 - q^2)   \right] \\
 &\qquad\qquad\qquad\qquad
 {} - q \mu_q^2 + \frac{q(1 + 2 q^2)}6   \,.
 }
This expression can easily be evaluated numerically for any $\mu_q^2$. Note that the same procedure gives $\mathcal{F}_0^{\infty}=0$, as required by conformal symmetry.

As mentioned above, the value of $\mu_q^2$ can be obtained from the saddle-point equation \eqref{Saddlemu}, which yields
 \es{solveaq}{
\sum_{j = q}^{\infty} \left( \frac{j +1/2}{\sqrt{ (j + 1/2)^2 + \mu_q^2 - q^2}} - 1 \right) -q= 0 \,.
 }
Note that one can obtain this equation directly by differentiating \eqref{omegad} with respect to $\mu_q^2$ without the need of zeta-function regularization.  Upon substituting the solution of \eqref{solveaq} into \eqref{FInfFinal}, one obtains the values of ${\cal F}_q^\infty$ given in Table~\ref{FLargeNTable}. 
\begin{table}[htp]
\begin{center}
\begin{tabular}{c|c|c}
$q$ & $\mu_q^2$ & ${\cal F}_q^\infty$ \\
\hline \hline
$0$ & $0$ & $0$ \\
$1/2$ & $-0.199806$ & $0.1245922$ \\
$1$ & $-0.397830$ & $0.3110952$ \\
$3/2$ & $-0.595457$ & $0.5440693$ \\
$2$ & $-0.792936$ & $0.8157878$ \\
$5/2$ & $-0.990344$ & $1.1214167$ 
\end{tabular}
\end{center}
\caption{A few values for the parameters $\mu_q^2$ and ${\cal F}_q^\infty$.}
\label{FLargeNTable}
\end{table}%
These values agree precisely with those obtained in \cite{Murthy:1989ps} by a very different method.\footnote{The quantity ${\cal F}_q^\infty$ should be identified with $2 \rho_{2q}$ in \cite{Murthy:1989ps}, while our $\mu_q^2$ should be identified with $-C_{2q}$ in \cite{Murthy:1989ps}.}

\section{$1/N_b$ corrections} \label{CORRECTIONS}
In this section we compute the next to leading oder correction to the dimensions of monopole operators. The systematics of the calculation are presented in the first four subsections, with the numerical results presented in Section \ref{NUMERICS}.
\subsection{Effective action at quadratic level}

To obtain the leading $1/N_b$ correction $\delta {\cal F}_q$ to the result of the previous section, one should consider the quadratic fluctuations of the gauge field and of the Lagrange multiplier around the saddle \eqref{AlambdaExpansion}.  Expanding \eqref{EffAction} at small $a_\mu$ and $\sigma$, one can write the quadratic term in the effective action as
 \es{EffectiveActionQuad}{
  {\cal S}^{(2)}_\text{eff} &= {\cal S}^{(2)}_{\sigma\sigma} + {\cal S}^{(2)}_{aa} + {\cal S}^{(2)}_{a\sigma} \,, \\
\mathcal{S}^{(2)}_{\sigma\sigma} &= 
\frac{N_b}{2} \int d^3 x d^3 x' \sqrt{g (x) } \sqrt{g (x') }  \sigma  (x)  D^q (x,x')  \sigma (x') \,, \\
\mathcal{S}^{(2)}_{aa} &= 
\frac{N_b}{2} \int d^3 x d^3 x' \sqrt{g (x) } \sqrt{g (x') }  a_\mu  (x) K^{q,\mu\mu'} (x,x') a_{\mu'} (x') \,, \\
 {\cal S}^{(2)}_{a\sigma} &= N_b
  \int  d^3 x d^3 x' \sqrt{g (x) } \sqrt{g (x') }  \sigma  (x) H^{q,\mu'} (x,x') a_{\mu'} (x') \,,
 }
where $D^q$, $K^{q}$, and $H^{q}$ are integration kernels whose expressions will be given shortly.  This effective action is non-local because it was obtained after integrating out the fields $\phi_\alpha$, which are massless.  The kernels appearing in \eqref{EffectiveActionQuad} can be written in terms of correlators of $\abs{\phi_\alpha}^2$ and of the current 
 \es{CurrentDef}{
   J_\mu = i \left[ \phi_\alpha^\ast (\nabla_\mu - i \mathcal{A}_\mu^q) \phi_\alpha - \phi_\alpha (\nabla_\mu + i \mathcal{A}^q_\mu) \phi_\alpha^\ast \right] \,,
 }
as
 \es{KernelsExplicit}{
  N_b D^q(x,x') &= \frac{1}{\mathfrak{g}^2} \langle \abs{\phi_\alpha(x)}^2 \abs{\phi_\alpha(x')}^2 \rangle_q  \,, \\
  N_b K^{q, \mu\mu'}(x,x') &= - \frac{1}{\mathfrak{g}^2} \langle J^\mu(x) J^{\mu'}(x') \rangle_q 
    + \frac{2}{\mathfrak{g}} g^{\mu\mu'} \delta(x - x')  \langle \abs{\phi_\alpha(x)}^2 \rangle_q \,, \\
  N_b H^{q, \mu'} (x,x') &= - \frac{i}{\mathfrak{g}^2} \langle \abs{\phi_\alpha(x)}^2   J^{\mu'}(x') \rangle_q \,,
 }
where the delta-function contains a factor of $1/\sqrt{g(x)}$ in its definition.  The correlators in \eqref{KernelsExplicit} are evaluated under the assumption that the gauge field and Lagrange multiplier are non-dynamical and fixed at their background values $A_\mu = {\cal A}_\mu^q$ and $i \lambda = \mu_q^2 $.  The subscript $q$ on the angle brackets in the expressions above serves as a reminder of these assumptions.

Performing the Gaussian integral over $a_\mu$ and $\sigma$, we can write the coefficient $\delta {\cal F}_q$ in \eqref{expand} as
 \es{GotDeltaFq}{
  \delta {\cal F}_q = \frac12\, \log \det{}' M^q \,, 
 }
where we defined the matrix of kernels
  \es{MDef}{
  M^q(x, x') \equiv \begin{pmatrix}
   D^q(x, x') & H^q_{\tau'}(x, x') & H^q_{i'}(x, x')  \\
   H^{q \tau}(x, x') & K^{q\tau}{}_{\tau'}(x, x') & K^{q\tau}{}_{i'}(x, x') \\
   H^{qi}(x, x') & K^{q\tau}{}_{i'}(x, x') & K^{qi}{}_{i'}(x, x') 
  \end{pmatrix} \,,
 }
with $i = \theta,\phi$ and the primed indices contracting with the index of the field at $x'$.  The prime on the determinant in \eqref{GotDeltaFq} means that when computing the functional determinant we should ignore the zero eigenvalues that are required to be present due to gauge invariance.\footnote{For a more detailed treatment of gauge fixing, see~\cite{Dyer:2013fja}.}  Our goal in the rest of this section is to calculate a regularized version of this determinant, thus obtaining $\delta {\cal F}_q$.

\subsection{Eigenvalues of the integration kernels}

One can start evaluating the expressions in \eqref{KernelsExplicit} in terms of the $N_b = \infty$ limit of the Green's function $G^q(x, x')$ for the complex scalars, which is defined by
 \es{GDef}{
  \langle \phi_\alpha(x) \phi^*_\beta(x') \rangle = \mathfrak{g} \, \delta_{\alpha \beta} G^q(x, x') \,.
 }
Performing the required Wick contractions in \eqref{KernelsExplicit}, we obtain
 \es{KernelsAgain}{
  D^q(x, x') &= G^q(x, x') G^{q*}(x, x')  \,, \\
  K^{q}_{\mu\mu'}(x, x') &= D_\mu G^q(x, x') D_{\mu'} G^{q\ast} (x, x') 
-  G^{q\ast} (x, x') D_\mu D_{\mu'} G^q(x, x')   \\
  &{}+ D_\mu G^{q*}(x, x') D_{\mu'} G^{q} (x, x') 
-  G^{q} (x, x') D_\mu D_{\mu'} G^{q*}(x, x') \\
  &{}+2 g_{\mu\nu} \delta(x-x') G^q(x, x) \,, \\
  H^{q}_{\mu'} (x, x') &= G^{q}(x, x') D_{\mu'} G^{q*}(x, x') - G^{q*}(x, x') D_{\mu'} G^{q}(x, x')  \,,
 }
where $D_\mu = \partial_\mu - i {\cal A}^q_\mu(x)$ and $D_{\mu'} = \partial_{\mu'} + i {\cal A}^q_\mu(x')$ denote the gauge-covariant derivatives in the presence of the background gauge field. 

In order to calculate the eigenvalues of the matrix of kernels \eqref{MDef} required for \eqref{GotDeltaFq}, we make use of the $S^2$ rotational symmetry and the translational symmetry along $\R$.  These symmetries imply that the eigenvectors of this matrix are of the form $e^{-i \omega \tau}$ times an appropriate (scalar or vector) spherical harmonic on $S^2$.  We will need the usual spherical harmonics $Y_{jm}(\theta, \phi)$, as well as the vector harmonics 
 \es{VectorHarm}{
\mathcal{X}_{i,j m} &= \frac{1}{\sqrt{j (j + 1)}} \partial_i Y_{j m} \,, \\
\mathcal{Y}^i_{j m} &= \frac{1}{\sqrt{j (j + 1)}} \frac{\epsilon^{ik}}{\sqrt{g}} \partial_k Y_{j m} \,,
 }
where $i,k=\theta,\phi$, and $\epsilon^{\theta\phi} = -\epsilon^{\phi\theta} = 1$. We can decompose the Lagrange multiplier fluctuation $\sigma$ and the gauge field fluctuation $a_\mu$ in terms of these modes. Because,  $\sigma$ is a scalar field, we only need the usual spherical harmonics for its mode expansion. $a_\tau$ is also decomposed in terms of $Y_{jm}$, while $a_i$ is decomposed using the vector harmonics $\mathcal{X}_{i,j m}$ and $\mathcal{Y}^i_{j m} $. We will refer to the former vector harmonics modes as $E$ modes and to the latter as $B$ modes, as they are the $S^2$ analogs of $E$ and $B$ modes familiar from other contexts: $E$ and $B$ modes have vanishing curl and divergence, respectively, and they transform like the $E$ and $B$ field under parity.

We can expand each of the kernels in Fourier modes as
\beq
D^q (x, x') = \int \frac{d \omega}{2 \pi} \sum_{j m} D^q_{j} (\omega) Y_{j m}  (\theta, \phi) Y^\ast_{j m} (\theta', \phi') 
e^{-i \omega (\tau - \tau')} \label{Dexp}
\eeq
 \es{Kexp}{
K^q_{\tau\tau} (x, x') &= \int \frac{d \omega}{2 \pi} \sum_{j m} K^{q,\tau\tau}_{j} (\omega) Y_{j m}  (\theta, \phi) Y^\ast_{j m} (\theta', \phi') e^{-i \omega (\tau - \tau')} \\
K^q_{ii'} (x, x') &= \int \frac{d \omega}{2 \pi} \sum_{j m}
   \Bigl[ K^{q,EE}_{j} (\omega)  \mathcal{X}_{i,j m}  (\theta, \phi) \mathcal{X}^{\ast}_{i',j m} (\theta', \phi')
    + K^{q,BB}_{j} (\omega) \mathcal{Y}_{i,j m}  (\theta, \phi) \mathcal{Y}^{\ast}_{i',j m} (\theta', \phi')\\
      &+ K^{q,EB}_{j} (\omega) \mathcal{X}_{i,j m}  (\theta, \phi) \mathcal{Y}^{\ast}_{i',j m} (\theta', \phi')
      + K^{q,EB*}_{j} (\omega) \mathcal{Y}_{i,j m}  (\theta, \phi) \mathcal{X}^{\ast}_{i',j m} (\theta', \phi') \Bigr]
         e^{-i \omega (\tau - \tau')} \\
K^q_{\tau i'} (x, x') &= \int \frac{d \omega}{2 \pi} \sum_{j m} \left[ K^{q,\tau E}_{j} (\omega) Y_{j m}  (\theta, \phi) {\cal X}^\ast_{i', j m} (\theta', \phi') + K^{q,\tau B}_{j} (\omega) Y_{j m}  (\theta, \phi) {\cal Y}^\ast_{i', j m} (\theta', \phi')\right] e^{-i \omega (\tau - \tau')} 
 }
 \es{Fexp}{
H^q_{\tau'} (x, x') &= \int \frac{d \omega}{2 \pi} \sum_{j m} H^{q,\tau}_{j} (\omega) Y_{j m}  (\theta, \phi) Y^\ast_{j m} (\theta', \phi') e^{-i \omega (\tau - \tau')} \\
H^q_{i'} (x, x') &= \int \frac{d \omega}{2 \pi} \sum_{j m} \left[ H^{q, E}_{j} (\omega)Y_{j m}  (\theta, \phi) {\cal X}^\ast_{i', j m} (\theta', \phi') + H^{q, B}_{j} (\omega)Y_{j m}  (\theta, \phi) {\cal Y}^\ast_{i', j m} (\theta', \phi') \right] e^{-i \omega (\tau - \tau')} 
 } 
and form the matrix of coefficients\footnote{We are indebted to Nathan Agmon whose work revealed that the original version of this equation contained a minus sign error \cite{AgmonThesis}.}  
\es{GotM}{
 {\bf M}^q_{j}(\omega) =    \begin{pmatrix}[c|ccc]
    D^{q}_j(\omega) & H^{q, B}_j(\omega)& H^{q, \tau}_j(\omega) & H^{q, E}_j(\omega)   \\ \hline
    -H^{q, B*}_j(\omega) & K^{q, B B}_j(\omega) & K^{q, \tau B}_j(\omega) & K^{q,  E B}_j(\omega)  \\
    -H^{q, \tau*}_j(\omega) & K^{q, \tau B*}_j(\omega) & K^{q, \tau\tau}_j(\omega) & K^{q, \tau E}_j(\omega)  \\ 
    -H^{q, E*}_j(\omega) & K^{q, E B*}_j(\omega) & K^{q, \tau E*}_j(\omega) & K^{q, E E}_j(\omega)  \\
   \end{pmatrix} \,.
}
Note that $S^2$ rotational symmetry implies that these coefficients do not depend on the quantum number $m$.   Note that the matrix $ {\bf M}^q_{j}(\omega)$ is not Hermitian, which can be traced to the fact that in the action~\eqref{CPNActionRescaled} the Lagrange multiplier field $\lambda$ appears multiplied by a factor of $i$.

The entries of this matrix are related by gauge invariance and $CP$ symmetry. Gauge invariance of the kernels in position space imply that at separated points\footnote{There are multiple equivalent ways to see that these equations are true. Using the definitions of the kernels in terms of correlators~\eqref{KernelsExplicit} they are the consequences of the Ward identity $\nabla^\mu J_\mu(x)=0$. Alternatively,~\eqref{EffectiveActionQuad} should be zero for a pure gauge configuration $a_\mu=\nabla_\mu\, \alpha(x)$, for arbitrary $\alpha(x)$. Partial integration readily gives~\eqref{GaugeInv}.} 
\es{GaugeInv}{
\nabla^\mu \, K^q_{\mu\mu'}(x,x')&=0\,, \qquad \nabla^{\mu'} \, K^q_{\mu\mu'}(x,x')=0\,,\\
\nabla^{\mu'} \, H^q_{\mu'}(x,x')&=0\,.
}
Plugging in the decompositions~\eqref{Kexp} and~\eqref{Fexp} in these conservation equations, we obtain that the Fourier space kernel ${\bf M}^q_{j}(\omega)$ should have the following eigenvectors with zero eigenvalue:
 \es{GaugeInvar}{
  \begin{pmatrix}
   0\,\big\vert& 0& i \omega & - \sqrt{j(j+1)}  
  \end{pmatrix}  {\bf M}^q_{j}(\omega) &= 0 \,, \qquad
    {\bf M}^q_{j}(\omega)
    \begin{pmatrix}
 0\,\big\vert & 0 &   -i \omega & -\sqrt{j(j+1)} 
    \end{pmatrix}^T = 0 \,,\\
 }
where the pure gauge eigenvector is written in $(\sigma\,\big\vert\,  B,\, \tau,\, E)$ components, just like ${\bf M}^q_{j}(\omega)$ in~\eqref{GotM}. From~\eqref{GaugeInvar} we can express $K^{q, \tau E}_j(\omega) $ and $K^{q, E E}_j(\omega) $ in terms of $K^{q, \tau \tau}_j(\omega) $.

The second restriction on the entries of ${\bf M}^q_{j}(\omega)$ comes from the $CP$ invariance of the theory and the monopole background around which we are working. Under $CP$ the modes of $\sigma$ and the $B$ modes of $a_\mu$ transform in the same way, while the $\tau$ and $E$ modes acquire a relative minus sign. Because the effective action~\eqref{EffectiveActionQuad} is invariant under $CP$, we conclude that there is no mixing between $\sigma,\, B$ and $\tau,\, E$ modes.

  These constraints imply that ${\bf M}^q_{j}(\omega)$ takes a block diagonal form. For $j>0$,
\es{GotMSimp}{
 {\bf M}^q_{j}(\omega) =    \begin{pmatrix}[c|ccc]
    D^{q}_j(\omega) & H^{q, B}_j(\omega) & 0 & 0  \\ \hline
    -H^{q, B*}_j(\omega) & K^{q, B B}_j(\omega) & 0 & 0  \\
    0 & 0 & K^{q, \tau\tau}_j(\omega) & \frac{-i \omega}{\sqrt{j(j+1)}} K^{q, \tau \tau}_j(\omega)  \\
    0 & 0 & \frac{i \omega}{\sqrt{j(j+1)}} K^{q, \tau \tau}_j(\omega) & \frac{\omega^2}{j(j+1)} K^{q, \tau\tau}_j(\omega)  \\
   \end{pmatrix} \,.
}
This matrix has eigenvalues:
\es{Mev}{
\lambda^q_{\pm}&=\frac{(D_{j}^{q}(\omega)+K_{j}^{q,BB}(\omega))\pm\sqrt{(D_{j}^{q}(\omega)-K_{j}^{q,BB}(\omega))^{2}-4|H_{j}^{q,B}(\omega)|^{2}}}{2} \,, \\
\lambda^q_{E}&=\frac{j(j+1)+\omega^{2}}{j(j+1)}K_{j}^{q,\tau\tau}\,,
}
as well as a zero eigenvalue corresponding to a pure gauge mode. 

When $j=0$, the harmonics ${\cal X}_{jm}$ and ${\cal Y}_{jm}$ are not defined, so the matrix ${\bf M}^q_{j}(\omega)$ reduces to the $2 \times 2$ matrix
 \es{GotMSimp0}{
 {\bf M}^q_{0}(\omega) =    \begin{pmatrix}[c|c]
    D^{q}_0(\omega) & 0  \\ \hline
    0 & K^{q, \tau\tau}_0(\omega) 
   \end{pmatrix} \,.
}
In addition, the only remaining vector harmonic $Y_{00}$ is a constant on $S^2$, and can be gauged away.  Thus gauge invariance imposes $K_0^{q, \tau\tau} ( \omega) = 0$, and the only non-vanishing eigenvalue is $D^{q}_0(\omega)$.

We will derive expressions for the entries of the matrices \eqref{GotMSimp}--\eqref{GotMSimp0} shortly.  After doing so, we can calculate $\delta {\cal F}_q$ from~\eqref{GotDeltaFq}. It is convenient to subtract $\delta {\cal F}_0 = 0$ from  $\delta \mathcal{F}_q $.\footnote{Note that as  the gauge fixing condition is independent of the monopole background, any possible contribution from the Faddeev-Popov ghosts cancels after subtracting the vacuum contribution.} The expression we would like to calculate becomes:
 \es{FqSubtracted}{
  \delta \mathcal{F}_q = \frac 12 \int \frac{d\omega}{2 \pi}\   \sum_{j=0}^{\infty}
 ( 2 j + 1) \log {\det{}' M^q\ov \det{}' M^0} \,.
 }
Using the expression for the eigenvalues from~\eqref{Mev} and that $H_{j}^{0,B}(\omega)=0$ by parity symmetry, then \eqref{FqSubtracted} becomes:
 \es{FqFinalSimp}{
  \delta \mathcal{F}_q = \frac 12 \int \frac{d\omega}{2 \pi}\left[\log \frac{D^q_0(\omega) }{ D^0_0(\omega)  } +   \sum_{j=1}^{\infty}
 ( 2 j + 1) \log \frac{K^{q, \tau\tau}_j(\omega) \left[D^q_j(\omega) K^{q, BB}_j(\omega) + \abs{H^{q, B}_j(\omega)}^2 \right]}{ D^0_j(\omega) K^{0, \tau\tau}_j(\omega) K^{0, BB}_j(\omega) } \right] \,.
 }

Explicit expressions for the coefficients in \eqref{GotMSimp} can be obtained by inverting \eqref{Dexp}--\eqref{Fexp}.   Let us explain how to do so for $D_j^q(\omega)$ first, and leave the details of how to perform analogous computations for the $K$ and $H$ kernels to Appendix~\ref{DETAILS}.  For $D_j^q(\omega)$ we obtain:
 \es{InvertD}{
  D_j^q(\omega) 2 \pi \delta(\omega - \omega') = \int d^3 x \, d^3 x'\, \sqrt{g(x)} \sqrt{g(x')} Y_{jm}^*(\theta, \phi) D^q(x, x') Y_{jm}(\theta', \phi') e^{i (\omega \tau - \omega' \tau')} \,.
 }
Since the LHS is independent of $m$, we can average the RHS over all possible values of $m$.  After performing the average, the RHS becomes invariant under performing a combined rotation in $(\theta, \phi)$ and $(\theta', \phi')$, so we can take the limit $\theta' \to 0$.  We can also use that $D^q(x, x')$ depends only on $\tau -\tau'$ to set $\omega' = \omega$ and remove the $\tau'$ integral.  The simplified expression is
 \es{InvertDAgain}{
  D_j^q(\omega) = \frac{4 \pi}{2j+1} \int d^3 x \, \sqrt{g(x)} \lim_{\substack{\theta' \to 0 \\ \tau' \to 0}} \sum_{m = -j}^j Y_{jm}^*(\theta, \phi) D^q(x, x') Y_{jm}(\theta', \phi') e^{i \omega \tau} \,.
 }
It is only the $m=0$ term that contributes to the sum.  Analogous formulas for the $K$ and $H$ kernels are given in \eqref{InvertAll}.  Using explicit formulas for the spherical harmonics, \eqref{InvertDAgain} can be simplified further to
 \es{InvertDSimp}{
  D_j^q(\omega)  =  \int d^3 x \, \sqrt{g(x)} P_j(\cos \theta)   D^q(x, 0) e^{i \omega \tau} \,,
 }
where by $x' = 0$ we mean the limit $\tau', \theta' \to 0$.  Similar expressions (albeit more complicated) can be obtained for the other coefficients appearing in \eqref{GotMSimp}--\eqref{GotMSimp0}.

\subsection{Kernels at $q=0$}

When $q=0$, one can obtain closed form formulas for the entries of the matrix \eqref{GotMSimp}--\eqref{GotMSimp0}.  In this case, the Green's function $G^0(x, x')$ can be obtained by conformally mapping the $\R^3$ one, namely $1/(4 \pi |x - x'|)$, and from $G^0$ one can construct position-space expressions for all the kernels in \eqref{KernelsAgain}.  The conformal mapping from flat space gives
 \es{G0}{
   G^0(x, x') = \frac{1}{4  \pi \sqrt{2( \cosh(\tau-\tau') - \cos \gamma )}} \,,
 }
where $\gamma$ is the angle between the 2 points on $S^2$:
\beq
\cos\gamma = \cos\theta \cos \theta^\prime + \sin \theta \sin \theta' \cos(\phi - \phi') \,.
\label{defgamma}
\eeq
Since $G^0(x, x')$ is real, eq.~\eqref{KernelsAgain} implies that $H^0(x, x') = 0$, and consequently $H^{0, B}_j(\omega) = 0$.

\subsubsection{The $D$ kernel at $q=0$}
\label{D0KERNEL}

Plugging \eqref{KernelsAgain} and \eqref{G0} into \eqref{InvertDSimp}, we obtain
 \es{Dj0First}{
   D_j^0 (\omega) = \frac{1}{8 \pi} \int_{-\infty}^{\infty} d  \tau \int_0^{\pi} \sin \theta d \theta
\frac{ e^{i \omega \tau} P_j (\cos\theta)}{2(\cosh\tau - \cos \theta) } \,.
 }
This  expression can be evaluated by first performing the $\theta$ integral, which yields
 \es{Dj0FirstAgain}{
   D_j^0 (\omega) = \frac{1}{8 \pi} \int_{-\infty}^{\infty} d  \tau e^{i \omega \tau} Q_j (\cosh \tau)\,,
 }
where $Q_j(x)$ is the Legendre function of the second kind. We can then expand the remaining integrand at large $\tau$,
 \es{GotDj0}{
  D_j^0(\omega) = \frac{1}{8 \pi} \int_{-\infty}^{\infty} d  \tau \sum_{n=0}^\infty \frac{(n+j)! \Gamma(n+1/2) }{n! \Gamma(n + j + 3/2)} e^{-(2n + j + 1)\abs{\tau}} e^{i \omega \tau} \,,
 }
 and perform the $\tau$ integral term by term.  The result can be written as \cite{Pufu:2013eda}
  \es{Dj0Final}{
   D_j^0 (\omega) = \abs{\frac{\Gamma( (j + 1 + i \omega)/2)}{4 \Gamma((j + 2 + i \omega)/2)}}^2 \,.
  }
Note that in deriving \eqref{Dj0Final} we encountered no divergences in the sums and integrals we performed.

\subsubsection{The $K$ kernels at $q=0$}

Next, we aim to find an expression for $K^{0,\tau\tau}_j(\omega)$. While the expression for $K^{0, \tau\tau}_j(\omega)$ that follows from \eqref{Kexp} is UV divergent  (as would be its flat space analog), the following difference is finite:
 \es{KttFirst}{
  K^{0, \tau\tau}_j(\omega) - K^{0, \tau\tau}_0(0) =  \frac{1}{8 \pi} \int_{-\infty}^{\infty} d  \tau \int_0^{\pi} \sin \theta d \theta
\frac{ 1 - \cos \theta \cosh \tau}{2(\cosh\tau - \cos \theta)^3 } \left[ e^{i \omega \tau} P_j (\cos\theta) -1 \right] \,.
 }
It can be checked that
 \es{LapIdentity}{
   \frac{ 1 - \cos \theta \cosh \tau}{2(\cosh\tau - \cos \theta)^3 }  = \nabla_{S^2}^2 \frac{-1}{4(\cosh\tau - \cos \theta) } \,.
 } 
Substituting \eqref{LapIdentity} into \eqref{KttFirst}, integrating by parts twice in the sphere directions, and using  $\nabla_{S^2}^2 P_j (\cos\theta) = -j(j+1) P_j(\cos \theta)$, one can easily show that 
 \es{KttFinal0}{
  K^{0, \tau\tau}_j(\omega) - K^{0, \tau\tau}_0(0) = \frac{j(j+1)}{2} D_j^0(\omega) \,.
 }

Similarly, it can be shown that
 \es{OtherKRelations}{
  K^{0,BB}_j (\omega) - K^{0,BB}_1 (0)  &= \frac{(\omega^2 + j^2)}{2} D^0_{j-1} (\omega) - \frac{1}{2} D^0_0 (0) \,,\\
  K^{0,EE}_{j} (\omega) - K^{0,EE}_1 (0) &= \frac{\omega^2}{2} D^0_j (\omega) \,, \\
  K^{0,\tau E}_j (\omega) &= -i \omega \frac{\sqrt{j (j + 1)}}{2} D^0_j (\omega) \,.
 }
These expressions are consistent with the requirements of gauge invariance in \eqref{GotMSimp}. They also agree with the flat-space limit expected 
at large $\omega$ and $j$, which was obtained in~\cite{Kaul:2008xw}. Also, it follows from~\eqref{GotMSimp0} that at $j=0$ gauge invariance requires $K^{0,\tau\tau}_0 (0) = 0$. From~\eqref{GotMSimp} we also know that $K^{0,EE}_1 (\omega)= \frac{\omega^2}{2} K^{0, \tau\tau}_1(\omega)$. Taking the $\omega\to 0$ limit implies $K^{0,EE}_1 (0) = 0$, as $K^{0, \tau\tau}_1(0)$ is finite.  Assuming these relations, we have
 \es{KFinal}{
   K^{0,\tau\tau}_{j} (\omega)  &= \frac{j (j + 1)}{2} D^0_j (\omega)  \,, \\
   K^{0,BB}_j (\omega)  &= \frac{(\omega^2 + j^2)}{2} D^0_{j-1} (\omega)   + C_0 \,, \\
   K^{0,EE}_{j} (\omega) &= \frac{\omega^2}{2} D^0_j (\omega) \,, \\
   K^{0,\tau E}_j (\omega) &=- i \omega \frac{\sqrt{j (j + 1)}}{2} D^0_j (\omega) \,,
 }
where the constant $C_0$ remains to be determined; we will see later around~\eqref{CqDef} that $C_0 = 0$. 

\subsection{Kernels for general $q$}

For general $q$, there is no simple closed form expression for the $\phi_\alpha$ Green's function.  One can determine an integral expression for it by first expanding the fields $\phi_\alpha$ in Fourier modes:
 \es{phiExpansion}{
  \phi_\alpha = \int \frac{d\omega}{2 \pi} \sum_{j = q}^\infty \sum_{m = -j}^j \Phi_{\alpha, jm}(\omega) Y_{q, jm}(\theta, \phi) e^{-i \omega \tau} \,,
 }
where $Y_{q, jm}$ are the monopole spherical harmonics introduced in \cite{Wu:1976ge,Wu:1977qk}.  At leading order in $N_b$, the action for the fields $\phi_\alpha$ becomes 
 \es{ActionLeading}{
   \mathcal{S}_\phi = \frac{1}{\mathfrak{g}} \sum_{j = q}^{\infty} \sum_{m=-j}^{j} \int \frac{d \omega}{2 \pi} \left[ \omega^2 + (j + 1/2)^2 + \mu_q^2 -q^2 \right]
         |\Phi_{\alpha, jm} ( \omega)|^2  \,,
 }
from which we can read off
 \es{PhiCorrelator}{
  \langle \Phi_{\alpha, jm}(\omega) \Phi_{\beta, j' m'}(\omega')^* \rangle &= 2 \pi \mathfrak{g} \delta(\omega - \omega') \delta_{\alpha\beta} \delta_{jj'} \delta_{mm'}  G_j(\omega) \,,
 }
with
 \es{GFourier}{
  G_j(\omega) &\equiv \frac{1}{\omega^2 + (j + 1/2)^2 + \mu_q^2 -q^2 } \,.
 }
From \eqref{GDef}, \eqref{PhiCorrelator}, \eqref{phiExpansion}, and \eqref{GFourier} we can write $G^q(x, x')$ as
 \es{Gq}{
   G^q (x, x') &= \sum_{j = q}^{\infty} \int \frac{d \omega}{2 \pi} e^{-i \omega(\tau - \tau')} \left[\sum_{m=-j}^j Y_{q,jm} (\theta, \phi) Y_{q,jm}^* (\theta', \phi') \right] 
  G_j(\omega) \\
   &= \sum_{j = q}^{\infty}  e^{-2 i q \Theta} F_{q,j} ( \gamma) \frac{e^{- E_{qj} |\tau - \tau'|}}{2 E_{qj}} \,,
 }
where in the second line we defined the polynomial in $\cos \gamma$ 
 \es{defF}{
F_{q,j} (\gamma) \equiv \sqrt{ \frac{2 j + 1}{4 \pi}}Y_{q,j(-q)} (\gamma,0) \,,
 }
and $\Theta$ is a phase factor discussed in~\cite{Wu:1977qk} that can be defined through
\beq
e^{i \Theta} \cos (\gamma /2) =\cos (\theta/2) \cos(\theta'/2) + e^{-i (\phi - \phi')} \sin (\theta/2) \sin(\theta'/2).
\label{defTheta}
\eeq
The angle $\gamma$ was defined in \eqref{defgamma}, and the energy $E_{q j}$ is
\es{Eqjfef}{
E_{qj} \equiv \sqrt{ (j+1/2)^2 + \mu_q^2 - q^2}.
}

\subsubsection{The $D$ kernel}

Let us first determine $D^q_j (\omega) $. Using \eqref{KernelsAgain}, \eqref{InvertDSimp}, and \eqref{Gq}, we have
 \es{GotDq}{
   D^q_j (\omega) = \frac{4 \pi }{2 j + 1} \sum_{j', j''=q}^{\infty} \int d^3 x \sqrt{g (x)} 
     F_{0,j} (\theta) F_{q,j'} (\theta) F_{q,j''} (\theta)  
     \frac{ e^{-(E_{q j'}  + E_{q j''})|\tau| + i \omega \tau}}{4 E_{qj} E_{q j'}} \,.
 }
Performing the $\tau$ integral, we can simplify this expression to
 \es{dql}{
D^q_j ( \omega ) &= \frac{8 \pi^2}{2 j + 1} \sum_{j', j''=q}^{\infty}\ \left[ \frac{ E_{qj'}  + E_{q j''} }
{
2  E_{qj'}  E_{qj''} ( \omega^2 + ( E_{qj'} +  E_{qj''})^2)
}
\right] \mathcal{I}_D (j,j',j'') \,,  
}
where
 \es{IDDef}{
   \mathcal{I}_{D} (j, j', j'') = \int_0^\pi \sin \theta d \theta F_{0,j} (\theta) F_{q,j'} (\theta) F_{q,j''} (\theta) \,.
 }
The $\theta$ integral can be performed analytically, and we have
 \es{ID}{
   \mathcal{I}_D (j, j', j'') = \left[ \frac{(2 j + 1) (2 j' + 1) (2 j'' + 1)}{32 \pi^3} \right]
   \begin{pmatrix}
    j & j' & j'' \\
    0 & -q & q
   \end{pmatrix}^2 \,.
 }
We can check that this result equals \eqref{Dj0Final} for $q=0$ and, for instance, for $j=0$
 \es{D00Check}{
   D_0^0 (\omega) &= \frac{1}{2 \pi} \sum_{j' = 0}^{\infty} \frac{1}{\omega^2 + (2 j' + 1)^2} = \frac{ \tanh(\pi \omega/2)}{8 \omega} \,.
 }
Note that the summation in \eqref{GotDq} is absolutely convergent.

\subsubsection{The $K$ and $H$ kernels}

Similarly, for the other kernels we can use \eqref{KernelsAgain}, \eqref{Kexp}, \eqref{Fexp}, and 
 \eqref{Gq} to obtain
  \es{kebl}{
      H^{q, \tau}_j(\omega) &= H^{q, E}_j(\omega) = K^{q, \tau B}_j(\omega) = K^{q, E B}_j(\omega)= 0  \,, \\   
    H^{q, B}_j(\omega) &=  \frac{16 q \pi^2 i}{(2 j + 1)\sqrt{j(j+1)}} 
       \sum_{j', j''=q}^{\infty}\ \left[ 
       \frac{ E_{qj'}  + E_{q j''} }
      {2  E_{qj'}  E_{qj''} ( \omega^2 + ( E_{qj'} +  E_{qj''})^2)}
       \right] \mathcal{I}_{H} (j,j',j'') \,, \\
    K^{q,\tau\tau}_j ( \omega )  &= \frac{8 \pi^2}{2 j + 1} \sum_{j', j''=q}^{\infty}
         \left[ \frac{-( E_{qj'}  + E_{q j''}) (\omega^2 + 4 E_{q j'} E_{q j''}) }
         {
         2  E_{qj'}  E_{qj''} ( \omega^2 + ( E_{qj'} +  E_{qj''})^2)
          }
          \right] \mathcal{I}_{D} (j, j', j'') +  \sum_{j' = q}^{\infty} \frac{(2 j' + 1)}{4 \pi E_{q j'}} \,, \\
    K^{q,\tau E}_j ( \omega ) &= \frac{8 \pi^2}{(2 j + 1)\sqrt{j ( j + 1) }} 
        \sum_{j', j''=q}^{\infty}\ \left[ \frac
         {i \omega ( E_{qj'}  - E_{q j''})}
        {2  E_{qj'}  E_{qj''} ( \omega^2 + ( E_{qj'} +  E_{qj''})^2)}
         \right] \mathcal{I}_\phi (j, j', j'') \,, \\
    K^{q,EE}_j ( \omega )  &= \frac{8 \pi^2}{(2 j + 1)j ( j + 1) } 
       \sum_{j',  j''=q}^{\infty}\ \left[ \frac
        { ( E_{qj'}  + E_{q j''}) }
        {2  E_{qj'}  E_{qj''} ( \omega^2 + ( E_{qj'} +  E_{qj''})^2)}
        \right] \mathcal{I}_E (j, j', j'')  
        +  \sum_{j' = q}^{\infty} \frac{(2 j' + 1)}{4 \pi E_{q j'}} \,, \\
    K^{q,BB}_j ( \omega )  &= \frac{8 \pi^2}{(2 j + 1)j ( j + 1) } 
        \sum_{j' , j''=q}^{\infty}\ \left[ \frac{ ( E_{qj'}  + E_{q j''}) }
         {2  E_{qj'}  E_{qj''} ( \omega^2 + ( E_{qj'} +  E_{qj''})^2)}
         \right] \mathcal{I}_B (j, j', j'')  
         +  \sum_{j' = q}^{\infty} \frac{(2 j' + 1)}{4 \pi E_{q j'}}  \,,       
 }
where
 \es{Iint}{
     {\cal I}_{H}(j, j', j'') &= \int_0^\pi d \theta\, 
           \sin \theta \tan \frac{\theta}{2} F_{0,j}^\prime (\theta) 
           F_{q,j'} (\theta) F_{q,j''} (\theta) \,, \\
     \mathcal{I}_\phi (j, j', j'') &= 
         \Bigl[ j'' (j'' + 1) - j' (j' + 1) \Bigr] \mathcal{I}_D ( j, j', j'') \,, \\
     \mathcal{I}_E (j, j', j'')  &= 
        -\Bigl[ j' (j' + 1) - j'' (j'' + 1) \Bigr]^2 \mathcal{I}_D ( j, j', j'') \,, \\
    \mathcal{I}_B (j, j', j'')  &= \int_0^\pi d\theta\, \sin \theta  
         \Biggl[\frac{4}{\sin \theta} F_{0,j}^\prime (\theta) 
         F_{q,j'}^\prime  (\theta) F_{q,j''}^\prime (\theta)
         - 4 q^2 \tan^2(\theta/2) F_{0,j}^{\prime\prime} (\theta) 
        F_{q,j'} (\theta) F_{q,j''} (\theta) \Biggr] \,.
 }
A detailed derivation of these formulas is contained in Appendix~\ref{DETAILS}.  The quantity ${\cal I}_D(j, j', j'')$ appearing in \eqref{Iint} was given explicitly in \eqref{IDDef}.  Similar explicit expressions for ${\cal I}_H(j, j', j'')$ and ${\cal I}_B(j, j', j'')$ are given in Appendix~\ref{app:IBIF}.

Note that while the expressions for $H^{q, \tau}_j(\omega) $, $H^{q, B}_j(\omega)$, and $K^{q, \tau E}_j(\omega)$ above are absolutely convergent, those for $K^{q, \tau\tau}_j(\omega)$, $K^{q, EE}_j(\omega)$, and $K^{q, BB}_j(\omega)$ are not and require regularization.  In order to regularize the latter, it is convenient to first compute the quantities
 \es{FiniteQuantities}{
  &K^{q, \tau\tau}_j(\omega) - K^{0, \tau\tau}_0(0) \,, \\
  &K^{q, EE}_j(\omega) - K^{0, EE}_1(0) \,, \\
  &K^{q, BB}_j(\omega) - K^{0, BB}_1(0) \,,
 }
which are free of divergences, and then add back the appropriately regularized values for $K^{0, \tau\tau}_0(0)$, $K^{0, EE}_1(0)$, and $K^{0, BB}_1(0)$.  As argued in the previous subsection, gauge invariance implies $K^{0, \tau\tau}_0(\omega) = K^{0, EE}_1(0) = 0$, but does not immediately determine $K^{0, BB}_1(0)$ denoted by $C_0$ in~\eqref{KFinal}.  We can return now to that issue.  Let us first examine $K^{0,\tau\tau}_0(\omega) = 0$.  From \eqref{kebl}, we have
 \es{Ktt00}{
  K^{0, \tau\tau}_0(0) &= \sum_{j'=0}^{\infty}
         \left[  \frac{-(2j' + 1) }
         {
         4 \pi E_{0j'} 
          }
          \right] +  \sum_{j' = 0}^{\infty} \frac{(2 j' + 1)}{4 \pi E_{0 j'}}  = 0 \,.
 }
(Each sum is divergent individually, but the combined summation is convergent.)  Next, we can examine $K^{0, EE}_1(0)$.  Using \eqref{kebl} and doing a bit of algebra, we have
\es{KEE10Again}{
  K^{0, EE}_1(0) &= \frac{1}{32 \pi } 
       \sum_{\abs{j' -  j''}=1} \left[ - 2 - \frac{j'' + 1/2}{j' + 1/2} - \frac{j' + 1/2}{j'' + 1/2} 
        \right]
        +  \sum_{j' = 0}^{\infty} \frac{(2 j' + 1)}{4 \pi (j' + 1/2)} \,.
 }
By using the symmetry between the summation in $j'$ and $j''$, this expression can be written further as
\es{KEE10Again2}{
  K^{0, EE}_1(0) &= \frac{1}{16 \pi } 
       \sum_{\abs{j' -  j''}=1} \left[ - 1 - \frac{j'' + 1/2}{j' + 1/2}  
        \right]
        +  \sum_{j' = 0}^{\infty} \frac{(2 j' + 1)}{4 \pi (j' + 1/2)} \,.
 }
Summing over $j''$ we can write this expression as
\es{KEE10Again3}{
  K^{0, EE}_1(0) &= -\frac{1}{4 \pi } 
       \sum_{j' = 0}^\infty \frac{ (j' + 1/2)}{j' + 1/2}
        +  \sum_{j' = 0}^{\infty} \frac{(2 j' + 1)}{4 \pi (j' + 1/2)} \,.
 }
Both of the individual sums in this expression as well as the combined summation are divergent, but gauge invariance dictates that $K^{0, EE}_1(0) = 0$.  This result should be thought of as a prescription.  It can also be justified in zeta-function regularization, in which \eqref{KEE10Again3} gives $K^{0, EE}_1(0) = \zeta(0, 1/2)/(4 \pi) = 0$.

Next, from \eqref{kebl}, we can also write an expression for $K^{0, BB}_1(0)$:  
 \es{KBB10}{
     K^{0,BB}_1 ( 0 )  &= \frac{1}{8\pi} 
        \sum_{j' =0}^{\infty}\ \left[ \frac{ - j' (j' + 1) (2j' + 1)}
         {(j' + 1/2)^3 }
         \right]
         +  \sum_{j' = 0}^{\infty} \frac{(2 j' + 1)}{4 \pi (j' + 1/2)}  \,.    
 }
Both terms are again divergent, but, using \eqref{KEE10Again3}, we can calculate
 \es{KBB10Subtracted}{
  K^{0,BB}_1 ( 0 )  - K^{0,EE}_1 ( 0 )  = \frac{1}{4\pi} \sum_{j' =0}^{\infty}  \frac{1}{(2 j' + 1)^2} = \frac{\pi}{32} = \frac 12 D_0^0(0) \,.
 }
Since $K^{0, EE}_1(0) = 0$, this equation proves that $C_0 = 0$ in \eqref{KFinal}. 

We can now provide alternate, appropriately regularized formulas for $K^{q, \tau\tau}_j(\omega)$, $K^{q, EE}_j(\omega)$, and $K^{q, BB}_j(\omega)$ that are manifestly convergent.  They are:
 \es{KAlt}{
    K^{q,\tau\tau}_j ( \omega )  &= \sum_{j'=q}^{\infty} \Biggl[ \frac{8 \pi^2}{2 j + 1}  \sum_{j''=q}^\infty
         \frac{-( E_{qj'}  + E_{q j''}) (\omega^2 + 4 E_{q j'} E_{q j''}) }
         {
         2  E_{qj'}  E_{qj''} ( \omega^2 + ( E_{qj'} +  E_{qj''})^2)
          }
          \mathcal{I}_{D} (j, j', j'') \\
          &\hspace{3.5in}{}+  \frac{(2 j' + 1)}{4 \pi E_{q j'}} \Biggr]  \,, \\
    K^{q,EE}_j ( \omega )  &= \sum_{j' = q}^\infty \Biggl[  \frac{8 \pi^2}{(2 j + 1)j ( j + 1) } 
       \sum_{j''=q}^{\infty}\  \frac
        { ( E_{qj'}  + E_{q j''}) }
        {2  E_{qj'}  E_{qj''} ( \omega^2 + ( E_{qj'} +  E_{qj''})^2)}
       \mathcal{I}_E (j, j', j'')   \\
        &\hspace{3.5in}{}+  \frac{(2 j' + 1)}{8 \pi E_{q j'}}  \Biggr] + C_q \,, \\
    K^{q,BB}_j ( \omega )  &= \sum_{j' = q}^\infty \Biggl[  \frac{8 \pi^2}{(2 j + 1)j ( j + 1) } 
       \sum_{j''=q}^{\infty}\  \frac
        { ( E_{qj'}  + E_{q j''}) }
        {2  E_{qj'}  E_{qj''} ( \omega^2 + ( E_{qj'} +  E_{qj''})^2)}
       \mathcal{I}_B (j, j', j'')   \\
        &\hspace{3.5in}{}+  \frac{(2 j' + 1)}{8 \pi E_{q j'}}  \Biggr] + C_q \,, \\    
 } 
where
 \es{CqDef}{
  C_q \equiv \sum_{j' = q}^\infty \frac{2j' + 1}{8 \pi E_{q j'}} - \sum_{j' = 0}^\infty \frac{2j' + 1}{8 \pi E_{0 j'}} 
   = \frac{1}{4 \pi} \left[ \sum_{j' = q}^{\infty} \left( \frac{j' +1/2}{\sqrt{(j' + 1/2)^2 + \mu_q^2 - q^2}} - 1 \right) - q \right] \,.
 }
The expression for $K^{q, \tau\tau}_j(\omega)$ was obtained by simply combining the two summations in the expression in \eqref{kebl}.  The expressions for $K^{q, EE}_j(\omega)$ and $K^{q, BB}_j(\omega)$ were obtained by subtracting $K^{q, EE}_1(0) = 0$ from the expressions in \eqref{kebl}.  In~\eqref{CqDef} we discover the saddle point equation for $\mu_q^2$~\eqref{solveaq} (obtained after tuning the coupling $\mathfrak{g}$ to the critical value $\mathfrak{g}_c$), thus $C_q = 0$.

One can check from \eqref{kebl} and \eqref{Iint} that the gauge-invariance relations \eqref{GotMSimp} are obeyed.  Such a check is most simply performed by matching the residues of the functions of $\omega$ at their poles at $\omega = \pm i (E_{q j'} + E_{q j''})$ for each $j'$ and $j''$, after symmetrization between $j'$ and $j''$---see Appendix~\ref{GAUGECHECK}.

Now that we have expressions \eqref{KAlt} we can begin evaluating the order $1/N_{b}$ corrections to the free energy, which is the subject of the next subsection.


\subsection{Numerical Results}\label{NUMERICS}

With the regularized formulas for the kernels in hand, we are almost ready to calculate the subleading correction $\delta {\cal F}_q$ to the free energy using \eqref{FqFinalSimp}.  Let us write this expression as
 \es{deltaFqNotation}{
  \delta {\cal F}_q = \frac 12 \int\frac{d\omega}{2 \pi} \sum_{j = 0}^\infty (2j+1) \, L_j^q(\omega) \,,
 }
where $L_j^q(\omega)$ can be read off from \eqref{FqFinalSimp}.  (See also \eqref{LDef}--\eqref{L0Def}.)  As shown in Appendix~\ref{sec:Asym}, at large $\omega$ and $j$ the integrand in this expression behaves as
 \es{LLargejom}{
  L_j^q(\omega) = \frac{8\mu_{q}^{2}}{\omega^{2}+(j+1/2)^{2}}+\ldots \,,
 }
thus rendering the integral \eqref{deltaFqNotation} linearly divergent.  There are several ways of understanding how to regularize this divergence.  One way is to use zeta-function regularization to write
 \es{ZetaReg}{
  \frac 12 \int\frac{d\omega}{2 \pi} \sum_{j = 0}^\infty (2j+1) \frac{8\mu_{q}^{2}}{\omega^{2}+(j+1/2)^{2}}
   = \frac 12 \sum_{j= 0}^\infty (2j+1) \frac{8\mu_{q}^{2}}{2j+1} = 4 \mu_q^2 \zeta(0, 1/2) = 0 \,.
 }
Then one can subtract \eqref{ZetaReg} from \eqref{deltaFqNotation}, and evaluate 
 \es{deltaFqNotationReg}{
  \delta {\cal F}_q = \frac 12 \int\frac{d\omega}{2 \pi} \sum_{j = 0}^\infty (2j+1) \, \left[ L_j^q(\omega)  - \frac{8 \mu_q^2}{\omega^2 + (j+1/2)^2} \right] 
 }
instead of \eqref{deltaFqNotation}.  This expression is no longer linearly divergent.

Another way of understanding the subtraction in \eqref{deltaFqNotationReg} is that the critical coupling $\mathfrak{g}_c$, which was obtained in \eqref{CouplingSaddleCrit} at leading order in $N_b$, receives $1/N_b$ corrections.  A similar phenomenon was encountered in \cite{Kaul:2008xw} when computing the thermal free energy at subleading order in $1/N_b$.  Just as in \cite{Kaul:2008xw}, it can be argued that
 \es{CouplingCritMoreExact}{
    \frac{4 \pi}{\mathfrak{g}_c} = \int \frac{d\omega}{2 \pi} \sum_{j=0}^\infty (2j+1) \frac{1}{\omega^2 + (j+1/2)^2} \left[ 1 + \frac{4}{N_b} \right]  + O(1/N_b^2)  \,.
 }
The $1/N_b$ term in this expression contributes to $\delta {\cal F}_q^\infty$ through the last term in \eqref{EffAction} precisely as the subtraction implemented in \eqref{deltaFqNotationReg}.   This expression can be derived rigorously by performing a careful renormalization analysis of theory on $S^2 \times \R$~\cite{ScalarQED}.

Even after the linear divergence in $\delta {\cal F}_q^\infty$ has been taken care of, this quantity is still potentially logarithmically divergent.   This logarithmic divergence cancels when using a regularization prescription consistent with conformal symmetry. In practice, we evaluate the integral in \eqref{deltaFqNotationReg} with a symmetric cutoff:
 \es{symmcut}{
 (j+1/2)^{2}+\omega^{2}<\Lambda^{2}\,.
 } 
This can be thought of as preserving rotational invariance on $\mathbb{R}^{3}$, as the high energy modes are insensitive to the curvature of the sphere. Given the kernels, (\ref{KAlt}), and the regularization described above, we are able to evaluate $\delta\mathcal{F}_{q}$ numerically.  To obtain good precision, we first evaluate \eqref{deltaFqNotationReg} numerically in a region $(j+1/2)^{2}+\omega^{2}<(\Lambda')^{2}$;  then in the region $(\Lambda')^2 < (j+1/2)^{2}+\omega^{2} < \Lambda^2$ we replace $L_j^q(\omega)$ in \eqref{deltaFqNotationReg} with the asymptotic expansion derived in Appendix~\ref{sec:Asym} (accurate up to terms of order $O(1/\left[(j + \frac 12)^2 + \omega^2 \right]^{7/2})$) and evaluate the integral analytically as $\Lambda \to \infty$.  We notice that the result converges very rapidly as we increase $\Lambda'$. (In practice, $\Lambda' = 10$ is already sufficiently large.)  Our results for $\delta\mathcal{F}_{q}$ are given in Table~\ref{FNLOTable}.   See also Table~\ref{tab:intro} where we collected the results from Tables~\ref{FLargeNTable} and~\ref{FNLOTable} together.

 \begin{table}[htp]
\begin{center}
\begin{tabular}{c|c}
$q$  & $\delta{\cal F}_q$ \\
\hline 
$0$ & $0$ \\
$1/2$ & $0.3815$   \\
$1$ &  $0.8745$ \\
$3/2$ &  $1.4646 $\\
$2$ & $2.1388$ \\
$5/2$ &  $2.8879$  \\
\end{tabular}
\end{center}
\caption{The coefficient $\delta{\cal F}_q$ in the large $N_b$ expansion \eqref{expand} of the ground state energy in the presence of $2q$ units of magnetic flux through $S^2$.}
\label{FNLOTable}
\end{table}%


\section{The large $q$ limit}\label{STABILITY}

Whereas the analytic computation of  ${\bf M}^q_{j}(\omega)$ seems to be a hopeless endeavor for finite $q$, we found that the $q\to\infty$ limit is tractable. The reason for the simplification is that this is essentially a flat space limit: reintroducing the radius $R$ of $S^2$ we have a strong magnetic field $B=q/R^2$ on the sphere at large $q$.  $\phi$ quanta move on Landau levels, which are localized on $1/\sqrt{B}= R/\sqrt{q}$ distances, hence they don't feel the effect of the curvature of the sphere. To leading order in $1/q$, the problem becomes the analysis of the $\CP^{N_b-1}$ in a constant magnetic field in flat space.

\begin{figure}[h!]
  \centering
    \includegraphics[width=0.5\textwidth]{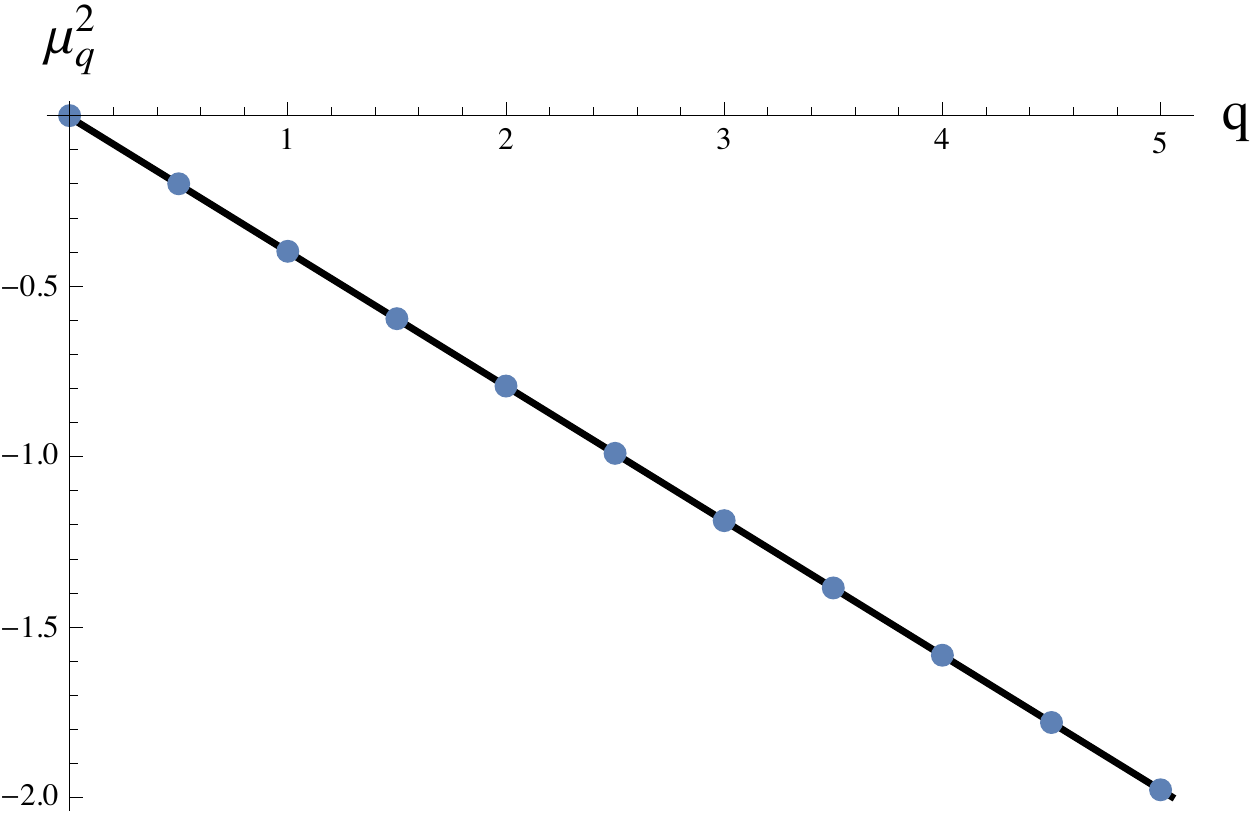}
      \caption{A comparison between the values of $\mu_q^2$ found by solving \eqref{solveaq} (blue points) and the large-$q$ analytical approximation \eqref{muExpNumerical} (solid black line).}
    \label{fig:muPlot}
\end{figure}

First, we want to calculate $\mathcal{F}_q^{\infty}$. We will be more cavalier about divergences than in the rest of the paper, and write, following~\eqref{FInfFinal},
 \es{FInfFinalFlat}{
\mathcal{F}_q^{\infty}
 &=  \sum_{j=q}^{\infty} (2j + 1)  
\left[(j + 1/2)^2 + \mu_q^2 - q^2 \right]^{1/2}\\
&= \sum_{n=0}^{\infty} ( 2q +2n+1)  
\left[2q(n + 1/2) + \mu_q^2 +(n+1/2)^2  \right]^{1/2}
   \,,
 }
where we introduced $n\equiv j-q$, and assumed zeta-function regularization as implicit. The resulting saddle point equation for $\mu_q^2$ is~\eqref{solveaq}:
 \es{solveaqFlat}{
\sum_{n = 0}^{\infty} \frac{2q +2n+1}{\sqrt{2q(n + 1/2) + \mu_q^2 +(n+1/2)^2}} = 0 \,.
 }
This equation has a solution only provided that we scale $\mu_q^2$ correctly with $q$, namely
\es{muExp}{
 \mu_q^2=2q\,\chi_0+\chi_1+O\le(1\ov q\ri)\,.
}
Plugging this Ansatz into~\eqref{solveaqFlat} and only keeping the leading terms, we obtain
 \es{solveaqFlat2}{
0=\sum_{n = 0}^{\infty} \frac{1}{\sqrt{(n + 1/2) + \chi_0}} =\zeta\le(\frac12,\, \frac12+\chi_0\ri)\,.
 }
Note that in obtaining this equation we assumed that $n\ll q$ even though we are summing over all positive $n$. This assumption is justified because the contribution of $n\gtrsim q$ is higher order in $1/q$. The constant $\chi_0$ can therefore be obtained as the root of the transcendental equation~\eqref{solveaqFlat2}. Going to one higher order we can determine $\chi_1$:
\es{chi1}{
\chi_1=-\chi_0^2+{3\, \zeta\le(-\frac12,\, \frac12+\chi_0\ri)\ov \zeta\le(\frac32,\, \frac12+\chi_0\ri)}\,,
}
which gives the large $q$ expansion for $\mu_q^2$:
\es{muExpNumerical}{
 \mu_q^2 \approx -0.39456\, q - 0.00456+O\le(1\ov q\ri)\,.
}
As seen from Figure~\ref{fig:muPlot}, \eqref{muExpNumerical} is in excellent agreement with the values of $\mu_q^2$ obtained from solving~\eqref{solveaq} at large $q$.\footnote{In \cite{Murthy:1989ps} it was noticed that $\mu_q^2 \approx - 2q/5$ at large $q$.} 

Plugging back into~\eqref{FInfFinalFlat}, we have that at leading order in $q$
\es{FInfFinalFlatAgain}{
\mathcal{F}_q^{\infty}
&= 2q   \sum_{n=0}^{\infty}  
\sqrt{2q(n + 1/2) +2q\,\chi_0 }=(2q)^{3/2}\, \zeta\le(-\frac12,\, \frac12+\chi_0\ri)\\
&=0.26408 \, q^{3/2}\,.
 }
Quite nicely, this equation can be understood in flat space terms: the Landau levels of a massive scalar field are given by 
\es{LandauLevel}{
E_n&=\sqrt{2B(n + 1/2)+m^2}={1\ov R}\,\sqrt{2q(n + 1/2) +2q\,\chi_0 }\,,
}
 and have degeneracy ${\cal N}={B\,\Vol(S^2)\ov 2\pi}=2q$, giving exactly~\eqref{FInfFinalFlat} for the free energy (if we set $R=1$). Note that the $q^{3/2}$ scaling of~\eqref{FInfFinalFlatAgain} follows from flat space dimensional analysis: the free energy density is an intensive quantity of mass dimension 3, hence it has to be independent of $R$, and we get ${\cal F}\sim B^{3/2}$.

While we leave a full evaluation of  ${\bf M}^q_{j}(\omega)$ at large $q$ to future work, as a first step we now derive large $q$ behavior of ${\bf M}^q_{j}(\omega)$ when $\omega=0$ and $j \ll q$.  This corresponds in the flat space limit to taking the momentum $p\ll BR$. At $\omega=0$, we only have to determine $D_j^q,\, H_j^{q,B},\, K_j^{q,BB}$, and $K_j^{q,\tau\tau}$, as the rest of the matrix elements vanish---see~\eqref{GotMSimp}. We can obtain a closed-form formula for these kernels by taking the explicit expressions for them, and expanding for large $q$. This is quite a tedious task, especially for $K_j^{q,BB}$, where we have to expand~\eqref{IBFull} for fixed $j,\,j',\, j''$ and large $q$. The resulting expressions can be summed over $j',\,j''$ analytically using zeta-function regularization. The results are given by the simple expression\footnote{For $j=0$ the matrix is $2\times2$, and the only nonzero element is $D^{q}_0(0)$, as in~\eqref{GotMSimp0}.}
\es{GotMSimpFlat}{
 {\bf M}^q_{j}(0) =  {\zeta\le(\frac32,\, \frac12+\chi_0\ri)\ov 8\, \pi\, \sqrt{2q}} \,  \begin{pmatrix}[c|ccc]
    \frac12 & i\sqrt{j(j+1)}\, \chi_0 & 0 & 0  \\ \hline
    i\sqrt{j(j+1)}\, \chi_0 & 2\,j(j+1)\, \chi_1 & 0 & 0  \\
    0 & 0 & 4\,j(j+1)\, (\chi_0^2+\chi_1) & 0  \\
    0 & 0 & 0 & 0 \\
   \end{pmatrix} \,.
}
One curious subtlety is that while the individual terms in $K_j^{q,BB},\, K_j^{q,\tau\tau}$ are $O(\sqrt{q})$, this $O(\sqrt{q})$ contribution vanishes upon summation over $j',\, j''$.  The subleading terms give the result in~\eqref{GotMSimpFlat}. This implies that we have to know the saddle point value of $\mu_q^2$ to first subleading order~\eqref{muExp} and $\chi_1$ makes appearance in the final result.

We obtain the flat space kernel by the replacement $j(j+1)\to p^2$ in~\eqref{GotMSimpFlat}. Our expression is valid for $p\ll BR$. It would be an interesting exercise to obtain the full quadratic effective action of the $\CP^{N_b-1}$ model in flat space in a constant magnetic field. 

\begin{figure}[h!]
  \centering
    \includegraphics[width=0.8\textwidth]{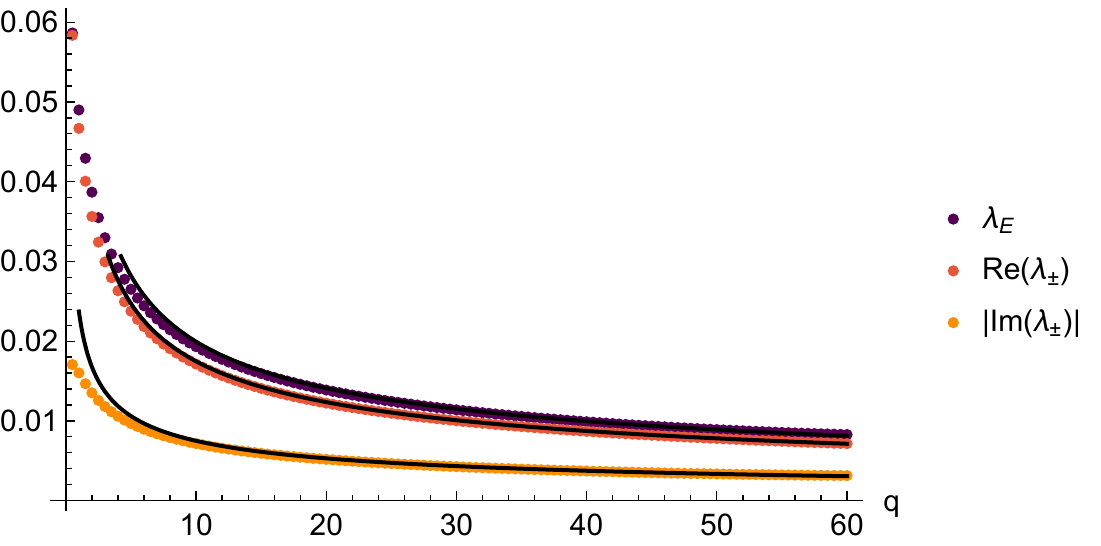}
      \caption{The numerical results for the three eigenvalues, $\lambda^{q}_{E}$, $\lambda^{q}_{+}$, and $\lambda^{q}_{-}$ are plotted against the analytic large $q$ value in black.}
    \label{fig:largeq}
\end{figure}

The matrix in \eqref{GotMSimpFlat} has one zero eigenvalue corresponding to the pure gauge mode.   The nonzero eigenvalues for $j=1$ are~\eqref{Mev}:
\es{MevFlat}{
\lambda^q_\pm&\approx -{0.055251\pm0.023717i\ov\sqrt{q}}\,, \qquad 
   \lambda^q_E\approx{0.063044\ov\sqrt{q}}\,.
}
We can check that \eqref{MevFlat} agrees with the numerical results---see Figure \ref{fig:largeq}.

\section{Conclusions}
\label{sec:conc}

We have presented here the leading correction to the large $N_b$ result for the scaling dimension of the monopole operator
in the $\CP^{N_b - 1}$ CFT\@. This correction was obtained by computing the Gaussian fluctuation determinant of the U(1) gauge field $A_\mu$,
and the Lagrange multiplier $\lambda$, on $S^2 \times \R$. Computation of higher order terms in the $1/N_b$ expansion appears to be possible by the present methods,
but will involve considerable effort.

Our computation now opens the possibility of quantitatively testing the most subtle and novel aspects of the theory of deconfined criticality \cite{SVBSF,SBSVF}
in two-dimensional lattice antiferromagnets. An important feature of this theory is the connection between the monopole operator
and the VBS operator of the antiferromagnet \cite{Read:1990zza,Read:1989zz,Murthy:1989ps}. This connection allows a Monte Carlo computation
of the monopole scaling dimension by measuring correlators of the VBS order in lattice models. We compared our present result with the
Monte Carlo studies and found very good agreement both in the dimension of the lowest monopole operator and in the number of scalar fields below which monopole operators become relevant; see Fig.~\ref{fig:qmc} and Table~\ref{tab:intro} in the Introduction.

We conclude by noting that
our present study is among the most complex theoretical computations of critical exponents which have been compared to Monte Carlo simulations.
Our calculation assumed conformal invariance to realize a framework in which the exponents could be determined, and it did not reduce to identifying poles
in a Feynman graph expansion \cite{zinn2002quantum}. 
It would be of great interest to apply the recent progress in bootstrap methods \cite{Rattazzi:2008pe, Simmons-Duffin:2015qma} 
to also determine such exponents.

\subsection*{Acknowledgments}

  We are very grateful to Nathan Agmon, whose joint work with SSP \cite{AgmonThesis} revealed a crucial typo in the first version of this paper. We thank Ribhu Kaul for valuable discussions and for providing the numerical results in Fig.~\ref{fig:qmc}.   We also thank Max Metlitski for useful discussions.  The work of ED was supported by the NSF under grant PHY-0756174. The work of MM was supported by the Princeton Center for Theoretical Science.  SSP was supported in part by the US NSF under grant No.~PHY-1418069.   The work of SS was supported by the NSF under grant DMR-1360789 and MURI grant W911NF-14-1-0003 from ARO. Research at Perimeter Institute is supported by the
Government of Canada through Industry Canada and by the Province of
Ontario through the Ministry of Research and Innovation.

\appendix


\section{Derivation of integration kernels}\label{DETAILS}

\subsection{Evaluation of Fourier transforms}

In this Appendix we explain how to obtain the explicit formulas appearing in \eqref{kebl}.  When dealing with vectors on $S^2$ it is convenient to use the frame $e^i_a$, with $i = \theta, \phi$ and $a = 1, 2$, defined by
 \es{FrameDefs}{
  e_1^i = (1, 0) \,, \qquad e_2^i = \left( 0, \frac{1}{\sin \theta} \right) \,.
 }
One can then convert the coordinate index of a quantity $v_i$ to a frame index by writing 
 \es{viToa}{
  v_a = e^i_a v_i \,.
 }
The frame index can then be raised and lowered with $\delta^{ab}$ and $\delta_{ab}$, respectively;  in other words, it makes no difference whether it is upper or lower.

Our starting point are equations ~\eqref{Dexp}--\eqref{Fexp}.  In the main text, we explained how \eqref{Dexp} can be inverted to obtain \eqref{InvertD}.  The analogous formulas for the other kernels are
 \es{InvertAll}{
  K^{q, \tau\tau}_j(\omega) &= \frac{4 \pi}{2 j + 1} \int d^3 x \lim_{x' \to 0}
   \sum_{m = -j}^j Y^*_{jm}(\theta, \phi) K_{\tau \tau'}^q(x, x') Y_{jm}(\theta', \phi') e^{i \omega \tau} \,, \\
  K^{q, \tau E}_j(\omega)  &= \frac{4 \pi}{2 j + 1} \int d^3 x \lim_{x' \to 0}
   \sum_{m = -j}^j Y^*_{jm}(\theta, \phi) K_{\tau a'}^q(x, x') {\cal X}_{jm}^{a'}(\theta', \phi') e^{i \omega \tau} \,, \\
  K^{q, E E}_j(\omega)  &= \frac{4 \pi}{2 j + 1} \int d^3 x \lim_{x' \to 0}
   \sum_{m = -j}^j {\cal X}^{a*}_{jm}(\theta, \phi) K_{a a'}^q(x, x') {\cal X}_{jm}^{a'}(\theta', \phi') e^{i \omega \tau} \,, \\
  K^{q, BB}_j(\omega)  &= \frac{4 \pi}{2 j + 1} \int d^3 x \lim_{x' \to 0}
   \sum_{m = -j}^j {\cal Y}^{a*}_{jm}(\theta, \phi) K_{a a'}^q(x, x') {\cal Y}_{jm}^{a'}(\theta', \phi') e^{i \omega \tau} \,, \\  
  H^{q, B}_j(\omega)  &= \frac{4 \pi}{2 j + 1} \int d^3 x \lim_{x' \to 0}
   \sum_{m = -j}^j Y^*_{jm}(\theta, \phi) H_{a'}^q(x, x') {\cal Y}_{jm}^{a'}(\theta', \phi') e^{i \omega \tau} \,, \\
  D^{q}_j(\omega)  &= \frac{4 \pi}{2 j + 1} \int d^3 x \lim_{x' \to 0}
   \sum_{m = -j}^j Y^*_{jm}(\theta, \phi) D^q(x, x') Y_{jm}(\theta', \phi') e^{i \omega \tau} \,. 
 }
 Using the spectral decomposition \eqref{Gq}, the position-space kernels appearing in this expression can be written as
 \es{KernelsExpanded}{
  D^q(x, x') &= \sum_{j', j''}  F_{q,j'} ( \gamma)  F_{q,j''} ( \gamma) \frac{e^{- (E_{qj'} + E_{qj''}) |\tau - \tau'|}}{4 E_{q j'} E_{qj''}}  \,, \\
  K^q_{\tau\tau'}(x, x') &= \sum_{j', j''}  F_{q,j'} ( \gamma)  F_{q,j''} ( \gamma) \frac{e^{- (E_{qj'} + E_{qj'')} |\tau - \tau'|}}{4 E_{q j'} E_{qj''}} \left[(E_{q j'} - E_{q j''})^2 - 2 (E_{q j'} + E_{q j''}) \delta(\tau) \right] \\
  &{}+ \sum_{j'} \frac{2j'+1}{4 \pi E_{qj'}} \delta(x - x') \,, \\
    K^q_{\tau a'} (x, x') &= \sum_{j', j''}   \frac{e^{- (E_{qj'} + E_{qj''}) |\tau - \tau'|}}{4 E_{q j'} E_{qj''}} (E_{qj'} -E_{qj''}) \sgn (\tau - \tau')  \\
  &\times \left( \left[ D_{a'}  e^{-2 i q \Theta} F_{q,j'} ( \gamma)   \right] \left[e^{-2 i q \Theta} F_{q,j''} ( \gamma) \right]^* -  \left[ e^{-2 i q \Theta} F_{q,j'} ( \gamma)   \right] \left[D_{a'} e^{-2 i q \Theta} F_{q,j''} ( \gamma) \right]^* \right) \,,  \\
  K^q_{a a'} (x, x') &= \sum_{j'} \frac{2j'+1}{4 \pi E_{qj'}} \delta^{a a'} \delta(x - x')  + \sum_{j', j''} \frac{e^{- (E_{qj'} + E_{qj''}) |\tau - \tau'|}}{4 E_{q j'} E_{qj''}} \\
  &\times \biggl( \left[ D_a e^{-2 i q \Theta} F_{q,j'} ( \gamma) \right] \left[D_{a'} e^{-2 i q \Theta} F_{q,j''} ( \gamma)  \right]^*
    + \left[ D_{a'} e^{-2 i q \Theta} F_{q,j'} ( \gamma)  \right] \left[D_a e^{-2 i q \Theta} F_{q,j''} ( \gamma) \right]^* \\
  &- \left[ e^{-2 i q \Theta} F_{q,j'} ( \gamma)   \right] \left[D_a D_{a'} e^{-2 i q \Theta} F_{q,j''} ( \gamma)  \right]^* 
   - \left[D_a D_{a'} e^{-2 i q \Theta} F_{q,j'} ( \gamma) \right] \left[e^{-2 i q \Theta} F_{q,j''} ( \gamma)  \right]^* \biggr) \,, \\
    H^q_{a'} (x, x') &= \sum_{j', j''} \frac{e^{- (E_{qj'} + E_{qj''}) |\tau - \tau'|}}{4 E_{q j'} E_{qj''}}  \\
  &\times \left( \left[  e^{-2 i q \Theta} F_{q,j'} ( \gamma)   \right] \left[D_{a'} e^{-2 i q \Theta} F_{q,j''} ( \gamma) \right]^* -  \left[D_{a'} e^{-2 i q \Theta} F_{q,j'} ( \gamma)   \right] \left[e^{-2 i q \Theta} F_{q,j''} ( \gamma) \right]^* \right) \,,
 }
where 
 \es{DDef}{
   D_a \equiv \partial_a - i {\cal A}_a^q (x)\,, \qquad
    D_{a'} \equiv \partial_{a'} + i {\cal A}_{a'}^q (x') \,.
 }

In taking the $x' \to 0$ limit in \eqref{InvertAll}, it is convenient to use the addition formula for the spherical harmonics
 \es{YAddition}{
  \sum_{m = -j}^j Y^*_{jm}(\theta, \phi) Y_{jm}(\theta', \phi') = F_{0j}(\gamma) \,, \qquad F_{0j}(\gamma) \equiv \frac{2j+1}{4 \pi} P_j(\cos \theta) \,,
 }
where $\gamma$ is the relative angle between the points $(\theta, \phi)$ and $(\theta', \phi')$ on $S^2$.  Taking derivatives of this formula and using the definition of the vector harmonics in \eqref{VectorHarm}, we have 
 \es{Limits}{
  \lim_{x' \to 0} \sum_{m = -j}^j Y^*_{jm}(\theta, \phi) Y_{jm}(\theta', \phi') &= F_{0j}(\theta) \,, \\
   \lim_{x' \to 0} \sum_{m = -j}^j Y^*_{j m}(\theta, \phi) {\cal X}^{a'}_{j m}(\theta', \phi') 
   &= \frac{1}{\sqrt{j(j+1)}} \begin{pmatrix}
      -F_{0j}'(\theta) & 0
     \end{pmatrix}  \begin{pmatrix}
      \cos \phi & \sin \phi \\
      -\sin \phi & \cos \phi
     \end{pmatrix} \,, \\
   \lim_{x' \to 0} \sum_{m = -j}^j Y^*_{j m}(\theta, \phi) {\cal Y}^{a'}_{j m}(\theta', \phi') 
   &= \frac{1}{\sqrt{j(j+1)}} \begin{pmatrix}
      0 & F_{0j}'(\theta)
     \end{pmatrix}  \begin{pmatrix}
      \cos \phi & \sin \phi \\
      -\sin \phi & \cos \phi
     \end{pmatrix} \,, \\     
   \lim_{x' \to 0} \sum_{m = -j}^j {\cal X}^{a*}_{j m}(\theta, \phi) Y_{j m}(\theta', \phi') 
   &= \frac{1}{\sqrt{j(j+1)}} \begin{pmatrix}
      F_{0j}'(\theta) \\ 0
     \end{pmatrix} \,, \\
   \lim_{x' \to 0} \sum_{m = -j}^j {\cal Y}^{a*}_{j m}(\theta, \phi) Y_{j m}(\theta', \phi') 
   &= \frac{1}{\sqrt{j(j+1)}} \begin{pmatrix}
      0 \\ -F_{0j}'(\theta) 
     \end{pmatrix} \,, \\     
  \lim_{x' \to 0} \sum_{m = -j}^j {\cal X}^{a*}_{j m}(\theta, \phi) {\cal X}^{a'}_{j m}(\theta', \phi') 
   &= \frac{1}{j(j+1)} \begin{pmatrix}
      -F_{0j}''(\theta) & 0\\ 
       0 & 
      -\frac{F_{0j}'(\theta)}{\sin \theta}  
     \end{pmatrix}  \begin{pmatrix}
      \cos \phi & \sin \phi \\
      -\sin \phi & \cos \phi
     \end{pmatrix} \,, \\
    \lim_{x' \to 0} \sum_{m = -j}^j {\cal Y}^{a*}_{j m}(\theta, \phi) {\cal Y}^{a'}_{j m}(\theta', \phi') 
   &= \frac{1}{j(j+1)} \begin{pmatrix}
      -\frac{F_{0j}'(\theta)}{\sin \theta} & 0\\ 
       0 & 
         -F_{0j}''(\theta)
     \end{pmatrix}  \begin{pmatrix}
      \cos \phi & \sin \phi \\
      -\sin \phi & \cos \phi
     \end{pmatrix}    \,.
 } 
We also have
 \es{UsefulFLimits}{
  \lim_{x' \to 0} \left[ e^{-2 i q \Theta(x, x')} F_{qj}(\gamma) \right] &= F_{qj}(\theta)  \,, \\
  \lim_{x' \to 0} D_a \left[ e^{-2 i q \Theta(x, x')} F_{qj}(\gamma) \right] &=  
    \begin{pmatrix}
      F_{qj}'(\theta) \\
      -i q \tan \frac{\theta}{2} F_{qj}(\theta) 
    \end{pmatrix} \,, \\
  \lim_{x' \to 0} D_{a'} \left[ e^{-2 i q \Theta(x, x')} F_{qj}(\gamma) \right] &=
   \begin{pmatrix}
    -F_{qj}'(\theta) & - i q \tan \frac{\theta}{2} F_{qj}(\theta)
   \end{pmatrix}  
   \begin{pmatrix}
    \cos \phi & \sin \phi \\
    -\sin \phi & \cos \phi 
   \end{pmatrix}  \,, \\
  \lim_{x' \to 0} D_a  D_{a'} \left[ e^{-2 i q \Theta(x, x')} F_{qj}(\gamma) \right] &=   
   \begin{pmatrix} 
    -F_{qj}''(\theta)  &  \frac{-iq (F_{qj}(\theta) + \sin \theta F_{qj}'(\theta))}{1 + \cos \theta} \\
    \frac{iq (F_{qj}(\theta) - \sin \theta F_{qj}'(\theta))}{1 + \cos \theta} & -\frac{F_{qj}'(\theta)}{\sin \theta} + q^2 \tan^2 \frac{\theta}{2} F_{qj}(\theta) 
    \end{pmatrix} \\
    &\qquad\qquad\qquad\qquad\qquad\qquad\times 
    \begin{pmatrix}
    \cos \phi & \sin \phi \\
    -\sin \phi & \cos \phi 
   \end{pmatrix}  \,. 
 }

Plugging in \eqref{Limits} and \eqref{UsefulFLimits} into \eqref{InvertAll} and performing the integrals over $\omega$ and $\phi$, we obtain \eqref{dql} and \eqref{kebl}, with
 \es{IDefsOriginal}{
  {\cal I}_D(j, j', j'') &\equiv \int_0^\pi d\theta\, \sin \theta F_{0j}(\theta) F_{qj'}(\theta) F_{qj''}(\theta)  \,, \\
  {\cal I}_H(j, j', j'') &\equiv \int_0^\pi d\theta\,  \sin \theta \tan \frac{\theta}{2} F_{0j}'(\theta) F_{qj'}(\theta) F_{qj''}(\theta) \,, \\
  {\cal I}_\phi(j, j', j'') &\equiv \int_0^\pi d\theta\, \sin \theta F_{0 j}'(\theta) 
  \left(F_{q j'}'(\theta) F_{q j''}(\theta) - F_{q j'}(\theta) F_{q j''}'(\theta) \right) \,, \\
  {\cal I}_E(j, j', j'') &\equiv   \int_0^\pi d\theta\, \sin \theta A_E(\theta) \,, \\
  {\cal I}_B(j, j', j'') &\equiv  \int_0^\pi d\theta\, \sin \theta A_B(\theta) \,,
 }
where
 \es{ADef}{
  A_E &= F_{0j}'' \left( 2 F_{qj'}' F_{qj''}' - F_{qj'} F_{qj''}'' - F_{qj'}'' F_{qj''}
   \right) - \frac{ F_{0j}'}{\sin^2\theta} \frac{\partial}{\partial \theta}  \left[F_{qj'} F_{qj''} \right] 
   -\frac{4q^2}{\sin \theta} \tan^2 \frac{\theta}{2} F_{0j}' F_{q j'} F_{qj''} \,, \\
  A_B &= \frac{1}{\sin \theta} \left[ -F_{0j}'' \frac{\partial}{\partial \theta}  \left[F_{qj'} F_{qj''} \right]  +F_{0j}' \left(2 F_{qj'}' F_{qj''}' - F_{qj'} F_{qj''}'' - F_{qj'}'' F_{qj''}
   \right) \right] \\
   &- 4 q^2 \tan^2 \frac{\theta}{2} F_{0j}'' F_{qj'} F_{qj''}  \,.  
 }

To simplify these expressions, we can use the fact that the monopole spherical harmonics $Y_{q, jm}(\theta, \phi)$ are eigenfunctions of the gauge-covariant laplacian on $S^2$ with eigenvalue $j(j+1) - q^2$:
 \es{MonEq}{
  (\nabla^\mu - i {\cal A}^{q \mu} ) (\nabla_\mu - i {\cal A}^q_\mu) Y_{q, j m}(\theta, \phi) = \left[ j(j+1) - q^2 \right] Y_{q, j m}(\theta, \phi) \,.
 }
Since $F_{qj}(\theta) = \sqrt{\frac{2 j + 1}{4 \pi}} Y_{q, j(-q)}(\theta, 0) $, we also have
 \es{FDerRelation}{
  \left[ \frac{1}{\sin \theta} \partial_\theta \left( \sin \theta \partial_\theta \right)
   - q^2 \tan^2 \frac{\theta}{2} + j(j+1) - q^2 \right] F_{qj}(\theta) = 0 \,.
 }

Eq.~\eqref{FDerRelation} implies that after integration by parts in the formula for ${\cal I}_\phi$ above, we have
 \es{IntParts}{
  {\cal I}_\phi(j, j', j'')
   = \int d\theta \sin \theta F_{0, j}(\theta) F_{q, j'}(\theta) F_{q, j''}(\theta)  \left[j'' (j'' + 1) - j' (j' + 1) \right] \,,
 }
thus obtaining the simplified form for ${\cal I}_\phi$ presented in \eqref{Iint}.

One can simplify ${\cal I}_E$ as follows.  From \eqref{FDerRelation}, it follows that
 \es{ThreeDers}{
  \frac{1}{\sin \theta} \partial_\theta \left( \sin \theta F_{qj'}'' \right)
   = F_{qj'}' \left[ - j'(j'+1) + \frac{1}{\sin^2 \theta} + \frac{2 q^2}{\sin \theta} \tan \frac{\theta}{2} \right] 
   + \frac{2q^2}{\sin \theta} \tan^2 \frac{\theta}{2} F_{qj'} 
 }
and similarly for $F_{qj''}$.  After integration by parts in the second and third terms in the first paranthesis in \eqref{ADef} one obtains
 \es{AESimp}{
  A_E &= 2 F_{0j}'' F_{qj'}' F_{qj''}' + F_{0j}' \left( F_{qj'}'' F_{qj''}' + F_{qj'}' F_{qj''}'' \right)
     +  \frac{q^2}{\cos^2 \frac{\theta}{2}} F_{0j}' \left( F_{qj'} F_{qj''} \right)'\\
      &- F_{0j}' \left[j'(j' + 1) F_{qj'}' F_{qj''} +  j''(j'' + 1) F_{qj'} F_{qj''}' \right] \,.
 }
Integrating by parts in the last term and using \eqref{FDerRelation}, one has
 \es{AESimp2}{
  A_E &=  2 F_{0j}'' F_{qj'}' F_{qj''}' + F_{0j}' \left( F_{qj'}'' F_{qj''}' + F_{qj'}' F_{qj''}'' \right)
     +\frac{q^2}{\cos^2 \frac{\theta}{2}}  F_{0j}' \left( F_{qj'} F_{qj''} \right)' \\
      &+ F_{0j}  F_{qj'}' F_{qj''}' \left[j'(j' + 1) +  j''(j'' + 1) \right]  \\
      &-  F_{0j}  F_{qj'} F_{qj''}\left[j'^2(j'+1)^2 + j''^2 (j''+1)^2 - \frac{q^2}{\cos^2 \frac{\theta}{2}} \left[ j'(j'+1) + j''(j''+1)  \right] \right] \,.
 }
Integrating by parts in the fourth term we get:
  \es{AESimp3}{
  A_E &=  2 F_{0j}'' F_{qj'}' F_{qj''}' + F_{0j}' \left( F_{qj'}'' F_{qj''}' + F_{qj'}' F_{qj''}'' \right)
     +\frac{q^2}{\cos^2 \frac{\theta}{2}}  F_{0j}' \left( F_{qj'} F_{qj''} \right)'\\
      &- F_{0j}' \left[ F_{qj'} F_{qj''}' j'(j' + 1) +   F_{qj'}' F_{qj''} j''(j'' + 1) \right] \\
      &-  F_{0j}  F_{qj'} F_{qj''}\left(j'(j'+1) - j'' (j''+1) \right)^2 \,.
 }
Combining the third and fourth terms using \eqref{FDerRelation} gives:
 \es{AESimp4}{
  A_E &=2 F_{0j}'' F_{qj'}' F_{qj''}' + F_{0j}' \left( F_{qj'}'' F_{qj''}' + F_{qj'}' F_{qj''}'' \right)
   + F_{0j}' F_{qj''}' \frac{1}{\sin\theta} \left( \sin \theta F_{qj'}' \right)'\\
   &+ F_{0j}' F_{qj'}' \frac{1}{\sin\theta} \left( \sin \theta F_{qj''}' \right)' 
   -  F_{0j}  F_{qj'} F_{qj''}\left(j'(j'+1) - j'' (j''+1) \right)^2 \,.
 }
Everything except for the last term is a total derivative, so
 \es{AESimp5}{
  A_E &= -  F_{0j}  F_{qj'} F_{qj''}\left(j'(j'+1) - j'' (j''+1) \right)^2 \,.
 }
The simplified form for ${\cal I}_E$ presented in \eqref{Iint} immediately follows.  The simplification of ${\cal I}_B$ is achieved through a similar integration by parts.

\subsection{Check of gauge invariance} \label{GAUGECHECK}

As discussed around~\eqref{GotMSimp}, gauge invariance implies:
  \es{GaugeInvarConditions}{
   i \omega K_j^{q, \tau\tau}(\omega) &= - \sqrt{j(j+1)} K_j^{q, \tau E} (\omega) \,, \\
   i \omega K_j^{q, \tau E}(\omega) &= \sqrt{j(j+1)} K_j^{q, E}(\omega) \,.
  }

We can check that the expression in \eqref{kebl} obey these relations by comparing residues at $\omega = \pm i (E_{q j'} + E_{q j''}) $.  We have
 \es{ResFraction}{
  \text{Res}_{\omega =  \pm i (E_{q j'} + E_{q j''})} \frac{1}{\omega^2 +  (E_{q j'} + E_{q j''})^2} = \mp \frac{i}{E_{q j'} + E_{q j''}} \,.
 }
 
From \eqref{kebl} and \eqref{Iint}, one can check that, for instance, when $j' \neq j''$,
 \es{res1}{
  \text{Res} \left(i\omega K_j^{q, \tau\tau}(\omega)\right) &=  i \frac{8 \pi^2}{2 j+1}
    \frac{(E_{q j'} - E_{q j''})  \left(E_{q j'}^2 - E_{q j''}^2\right)}
    {2 E_{q j'} E_{q j''}} {\cal I}_D(j, j', j'') \\
    &= -\text{Res} \left(\sqrt{j(j+1)} K_j^{q, \tau E}(\omega) \right) \,,
 }
so the first gauge invariance relation is verified.  In checking this relation it was important that ${\cal I}_\phi(j, j', j'') = (E_{qj''}^2 - E_{qj'}^2) {\cal I}_D(j, j', j'')$.  A similar check shows that the second gauge invariance condition in \eqref{GaugeInvarConditions} is satisfied.


\section{Formulas for ${\cal I}_B$ and ${\cal I}_H$} \label{app:IBIF}

\subsection{Results for $q=0,\,1/2$}

We have only been able to perform the integrals as written in~\eqref{Iint} for $q=0,\,1/2$. We had to use a different method to obtain results for arbitrary $q$; the details are given in the next subsection. We give the results below.

For $q = 0$, we have
 \es{IB0}{
  {\cal I}_B(j, j', j'') = -\frac{(2j + 1) (2 j' + 1) (2 j''+1)(j' + j'' - j)(j + j' - j'') (j + j'' - j')}
    {32 \pi^3} \\
   \times \left[j + j' + j'' - \frac{1}{j + j' + j''} \right]  
   \begin{pmatrix}
    j-1 & j' - 1 & j'' - 1 \\
    0 & 0 & 0
   \end{pmatrix}^2 \,.
 } 

For $q=1/2$, we have
 \es{I1}{
    {\cal I}_H (j, j', j'') &= {\cal I}_D (j, j', j'') \times 
     \begin{cases}
      j(j+1) - (j' - j'')^2 & \text{for $j + j' + j''$ odd} \\
      j(j + 1) - (j' + j'' + 1)^2 & \text{for $j + j' + j''$ even}
     \end{cases} \\
    {\cal I}_B ( j, j' , j'' ) &= -\mathcal{I}_D ( j, j' , j'' ) \times
 \begin{cases}
    \Bigl[ (j' - j'')^2 - j ( j + 1) \Bigr]^2 &  \mbox{for $j + j' + j''$ odd} \\
    \Bigl[ (j' + j'' + 1)^2 - j ( j + 1) \Bigr]^2 & \mbox{for $j + j' + j''$ even}
 \end{cases}
 }

\subsection{Results for arbitrary $q$}

For arbitrary $q$ we were not able to perform the integrals in~\eqref{Iint} directly. Rather, we employed a method developed in~\cite{Dyer:2013fja} to obtain expressions for the kernels. From these expressions we were able to read off explicit formulas for ${\cal I}_B$ and ${\cal I}_H$. We briefly introduce the method below, and give the end results, which are somewhat complicated. 

In this subsection we will use a different basis for monopole vector harmonics than the one used in the main text.  We denote the harmonics in the new basis by $U_{q,jm}^\mu,\, V_{q,jm}^\mu,\, W_{q,jm}^\mu$.  For completeness we provide the details needed for the calculation, but we do not elaborate on the technique. For a detailed discussion see~\cite{Dyer:2013fja}. 

The vector spherical harmonics that we need are tangent vectors to $S^{2}\times\mathbb{R}$. There are a few natural choices of basis for this tangent space. The basis we used in the body of the text is just the basis induced from the round coordinates on $S^{2}\times\mathbb{R}$, and is given explicitly in~\eqref{FrameDefs}.

There is also a cartesian basis given by conformally mapping the natural frame basis of $\R^3$ to $S^2\times \R$. We write the standard line element on $\R^3$ in spherical coordinates as
\es{R3Metric}{
   ds_{\R^3}^2 &= d\vec{x}^2= e^{2\tau}\,\le[d\tau^2+ d\theta^2 + \sin^2 \theta d \phi^2 \ri] \qquad  \vec{x} \equiv   e^\tau \begin{pmatrix}
     \sin \theta \cos \phi & \sin \theta \sin \phi & \cos \theta 
    \end{pmatrix} \,.
    }
The metric on $S^2\times \R$ is obtained by rescaling the $\R^3$ metric \eqref{R3Metric} by $e^{-2\tau}$:
\es{RTimesS2}{
ds_{S^2\times \R}^2=d\theta^2+\sin^2\theta\, d\phi^2+d\tau^2 \ .
}
We obtain the frame on $S^2 \times \R$ by the conformal transformation of the standard frame $e^a_{\mathbb{R}^{3}}=dx^a$ on $\R^3$:
\es{Frame}{
 e^a = e^{-\tau} dx^a \ .
}
Explicitly,
\es{veilmat}{
e^{a}_{\mu}&=\left(
\begin{array}{ccc}
 \cos (\phi ) \sin (\theta ) & \cos (\theta ) \cos (\phi ) & -\sin (\theta ) \sin (\phi ) \\
 \sin (\theta ) \sin (\phi ) & \cos (\theta ) \sin (\phi ) & \cos (\phi ) \sin (\theta ) \\
 \cos (\theta ) & -\sin (\theta ) & 0 \\
\end{array}
\right)
}
The frame satisfies the relation
\es{cartFrame}{
 e^a_{\mu}e^{b}_{\nu}g^{\mu\nu} = \delta^{ab}\,.
}
There is yet another basis for the tangent space, in which the angular momentum operators, $\vec{J}^{2}$, and $J_{z}$ are diagonal. This basis is given by the vielbeins:
\es{angularp}{
e^{\pm}&=\frac{1}{\sqrt{2}}e^{-\tau}(\mp dx+ idy)\\
e^{z}&=e^{-\tau}dz\,,
}
with $x$, $y$, $z$ coordinates on $\mathbb{R}^{3}$.

In this raising and lowering basis the expressions of the vector spherical harmonics for $j>q$ are:
 \es{S1Harm}{
  U^{s}_{q, \, j m}(\hat{n}) = 
 \begin{pmatrix}
\sqrt{\frac{(j-m+1) (j-m+2)}{(2j+2) (2 j+3)}}Y_{q, j+1,m-1}(\hat{n})\\
-\sqrt{\frac{(j-m+1) (j+m+1)}{(j+1) (2 j+3)}}Y_{q, j+1,m}(\hat{n})\\
\sqrt{\frac{(j+m+1) (j+m+2)}{(2j+2) (2 j+3)}}Y_{q, j+1,m+1}(\hat{n})
 \end{pmatrix}\,,\\
 V^{s}_{q, \, j m}(\hat{n}) = \begin{pmatrix}
-\sqrt{\frac{(j-m+1) (j+m)}{2j (j+1)}}Y_{q, j,m-1}(\hat{n})\\
\frac{m}{\sqrt{j(1+j)}}Y_{q, j m}(\hat{n})\\
\sqrt{\frac{(j-m) (j+m+1)}{2j (j+1)}}Y_{q, j,m+1}(\hat{n})
 \end{pmatrix} \,, \\
W^{s}_{q, \, j m}(\hat{n}) = 
 \begin{pmatrix}
\sqrt{\frac{(j+m-1) (j+m)}{2j (2 j-1)}}Y_{q, j-1,m-1}(\hat{n})\\
\sqrt{\frac{(j-m) (j+m)}{j (2 j-1)}}Y_{q, j-1,m}(\hat{n})\\
\sqrt{\frac{(j-m-1) (j-m)}{2j (2 j-1)}}Y_{q, j-1,m+1}(\hat{n})
 \end{pmatrix}\,.
 }
Here, $U,V,W$ all have total angular momentum $j$, and $s$ runs over $\{+,z,-\}$. 

If $j=q$, we only have two modes  $U^{s}_{q, \, j m}(\hat{n})$ and $V^{s}_{q \, j m}(\hat{n})$. If $j=q=0$, only the $U^{s}_{q,\, j m}(\hat{n})$ mode is non vanishing. For $q\geq1$, if we have $j=q-1$ only the mode $U_{q,\, j m}(\hat{n})$ is non-zero. We don't encounter this case in this paper. From the raising and lowering basis we can go to the coordinate basis by contracting these expressions with $e_s^\mu$ in~\eqref{angularp}. Then 
$U_{q,jm}^\mu,\, V_{q,jm}^\mu,\, W_{q,jm}^\mu$  are themselves vectors of ordinary monopole harmonics (multiplied by some coordinate dependent factors coming from the $e_s^\mu$).

A key formula that we will use is
\es{YDerivative}{
D^\mu \le(Y_{q,jm}(\hat n)e^{-i\omega \tau}\ri)=& \begin{pmatrix}
      (j +i\omega) \sqrt{{(j + 1)^2 - q^2\ov(j + 1) (2 j + 1)}} \quad
     -\frac{q (1 -i\omega)}{\sqrt{j (j + 1)}} \quad
      (j + 1 - i\omega) \sqrt{\frac{j^2 - q^2}{ j (2 j + 1)}}
         \end{pmatrix} 
         \begin{pmatrix}
     U_{q,jm}^\mu\, e^{-i\omega \tau}\\
      V_{q,jm}^\mu\, e^{-i\omega \tau}\\ 
      W_{q,jm}^\mu\, e^{-i\omega \tau}
      \end{pmatrix}  \,,    
}
where we have to take the scalar product with the coefficient vector. 

After these preparatory steps we can present our method. We want to calculate the matrix elements of the kernel, ${\bf K}^{q}_j(\omega)$, but now in the basis given by \eqref{S1Harm}. To do this we have to invert~\eqref{Kexp}. As explained below~\eqref{InvertD}, using rotational symmetry we can average over the quantum number $m$ to get a formula that is easier to treat:
\es{FourierKernelAverage}{
 \le[{\bf K}^{q}_j(\omega)\ri]_{XZ} &=\frac{1}{2 j+1}\,\sum_{m=-j}^j\, \int d^3xd^3x'\,\sqrt{g(x)}\,\sqrt{g(x')}\, X^{\mu*}_{jm}(\hat n)\, {\cal K}^q_{ \mu\mu'}(x,x') \, Z^{\mu'}_{jm}(\hat n')\, e^{i\omega(\tau-\tau')}\,,
}
where $X,Z\in\{U,V,W\}$.

${\cal K}^q_{ \mu\mu'}(x,x')$ is given in terms of Green's functions in~\eqref{KernelsAgain}.
We use the first line of~\eqref{Gq} to do the Fourier decomposition of $G^q(x,x')$ instead of the second line, which we used in the rest of the paper. The benefit of treating $x$ and $x'$ separately, rather than introducing the relative angle, $\gamma$ is an algorithmic method to evaluate the kernels at the expense of having long formulas. 

Plugging into~\eqref{KernelsAgain} we get
\es{KSchematic}{
 {\cal K}^q_{ \mu\mu'}(x,x')=&\sum_{\substack{j',m'\\ j'',m''}}\int {d\omega'\ov 2\pi}\,{d\omega''\ov 2\pi}\ G_{j'}(\omega')\,G_{j''}(\omega'')\,\le[D_\mu \le(Y_{q,j'm'}(\hat n)e^{-i\omega'\tau} \ri) Y^*_{q,j''m''}(\hat n)e^{i\omega''\tau}\ri]\\
& \times \le[Y^*_{q,j'm'}(\hat n')e^{i\omega'\tau'} D_{\mu'} \le(Y_{q,j''m''}(\hat n')e^{-i\omega''\tau'} \ri) \ri]+\text{other distribution of derivatives}\,,
}
where we only wrote down explicitly the term coming from the first term in~\eqref{KernelsAgain}, and we grouped spherical harmonics at the same point inside square brackets. For the derivative of the spherical harmonics we can use~\eqref{YDerivative}, and plugging this into~\eqref{FourierKernelAverage} we get an expression involving one scalar and two vector spherical harmonics at the two spacetime points $x,x'$. We can then use~\eqref{S1Harm} to reduce the whole expression to the sum of products involving six ordinary spherical harmonics, three at each spacetime point $x,x'$.\footnote{It is more economical to use the frame basis~\eqref{S1Harm} to contract vector indices between the vector harmonics.} As anticipated this is a tedious, but algorithmic task.

We can actually perform the integral over $x'$ with no work. Because we averaged over $m$ in~\eqref{FourierKernelAverage},  the integrand depends only on the relative angle between $\hat n$ and $\hat n'$.  The integral with respect to $\hat n$ is therefore independent of $\hat n'$, so we can choose $\hat{n}^{\prime}$ to point in the $\hat{z}$ direction and replace the integral with respect to $\hat n'$ by a  factor of $4 \pi$.  Using
\begin{eqnarray}
Y_{q,\ell m}(\hat{z})&=&\delta_{q,-m}\sqrt{\frac{2\ell+1}{4\pi}} \,,
\end{eqnarray}
we get rid of three monopole harmonics.   The remaining angular integral over the product of three harmonics can be evaluated using some properties of monopole harmonics \cite{Wu:1977qk}.
 \es{YConj}{
   Y_{q, \ell m} (\hat n)^* = (-1)^{q+m}Y_{-q, \ell,-m} (\hat n) \,,
 }
and
 \es{ThreeJ}{
\int d\hat{n}\,&Y_{q,\ell m}(\hat{n}) Y_{q^{\prime},\ell^{\prime}m^{\prime}}(\hat{n})Y_{q^{\prime\prime},\ell^{\prime\prime}m^{\prime\prime}}(\hat{n}) \\
  &=(-1)^{\ell+\ell^{\prime}+\ell^{\prime\prime}}\sqrt{\frac{(2\ell+1)(2\ell^{\prime}+1)(2\ell^{\prime\prime}+1)}{4\pi}}
  \begin{pmatrix}
   \ell & \ell' & \ell'' \\
   q & q' & q''
  \end{pmatrix}
    \begin{pmatrix}
   \ell & \ell' & \ell'' \\
   m & m' & m''
  \end{pmatrix} \,,
 }
where $\begin{pmatrix}
 j & j' & j'' \\
 m & m' & m''
\end{pmatrix}$ is the Wigner 3$j$-symbol. 

After preforming the frequency integrals over $\omega',\, \omega''$, we get the answer as a sum of a large number of 3$j$-symbols, as many of the angular momentum indices on $Y_{q,jm}$ are shifted in~\eqref{S1Harm}. We can use some identities to reduce the number of 3$j$-symbols that appear in our final answer, but we are still left with long expressions. The same logic also gives $\le[{\bf H}^{q}_j(\omega)\ri]_{X}$, and ${ D}^{q}_{j}(\omega)$, though the latter does not involve any vector harmonics.

There is an easy translation between the integrals and the kernels~\eqref{kebl}. Because $V_{0,jm}$ is identical to the $B$-mode of the gauge field, we can read off ${\cal I}_H$ and ${\cal I}_B$ from $\le[{\bf H}^{q}_j(\omega)\ri]_{V}$ and $\le[{\bf K}^{q}_j(\omega)\ri]_{VV}$.  To simplify the expressions somewhat, we introduce the following notation for the product of two $3j$-symbol that will appear in our formulas
\es{Mon3j}{
\left[
\begin{array}{ccc}
 j & j'' & j' \\
 0 & m & -m \\
 0 & n & -n \\
\end{array}
\right]
\equiv\left(
\begin{array}{ccc}
 j & j'' & j' \\
 0 & m & -m \\
\end{array}
\right) 
\times \left(
\begin{array}{ccc}
 j & j'' & j' \\
 0 & -n & n \\
\end{array}
\right)
}

With this notation, we have:
\es{IFFull}{
&{\cal I}_H ( j, j' , j'' ) ={1\ov 64\,  q\,  \pi^3} \times\\
&\le\{
\resizebox{\textwidth}{!}{\begin{tabular}{ccc}
$  -(2 j+1) \left(2 j'+1\right) \left(2 j''+1\right) \
\sqrt{\left(j'-q+1\right) \left(j'+q\right) \left(j''-q+1\right) \
\left(j''+q\right)}$ & $ \times$ & $\left[
\begin{array}{ccc}
 j & j'' & j' \\
 0 & -q & q \\
 0 & q-1 & 1-q \\
\end{array}
\right]$ \\ 
 $  -2 (2 j+1) q \left(2 j'+1\right) \left(2 j''+1\right)$ & $ \times$ & $\left[
\begin{array}{ccc}
 j & j'' & j' \\
 0 & -q & q \\
 0 & q & -q \\
\end{array}
\right]$ \\ 
 $ + (2 j+1) \left(2 j'+1\right) \left(2 j''+1\right) \
\sqrt{\left(j'-q\right) \left(j'+q+1\right) \left(j''-q\right) \
\left(j''+q+1\right)}$ & $ \times$ & $\left[
\begin{array}{ccc}
 j & j'' & j' \\
 0 & -q & q \\
 0 & q+1 & -q-1 \\
\end{array}
\right]$
\end{tabular}}
\ri\}
}

\es{IBFull}{
&{\cal I}_B ( j, j' , j'' ) ={1\ov 128 q j' (j' + 1) j'' ( j''+1) \pi^3} \times\\
&\le\{
\resizebox{\textwidth}{!}{\begin{tabular}{ccc}
$ \left(
\begin{array}{c}
 -(2 j+1) j' \left(j'+1\right) \left(2 j'+1\right) \left(j''+1\right)^2 \left(q+j''\right) \\
 \times \sqrt{\left(-q^2+q+j'^2+j'\right) \
\left(-q^2+q+j''^2-j''\right)} \left(-j^2-j+j'^2+j'+j'' \left(-2 q+j''-1\right)\right) \\
\end{array}
\right)$ & $ \times$ & $\left[
\begin{array}{ccc}
 j & j''-1 & j' \\
 0 & -q & q \\
 0 & q-1 & 1-q \\
\end{array}
\right]$ \\ 
 $ + \left(
\begin{array}{c}
 -2 (2 j+1) j' \left(j'+1\right) \left(2 j'+1\right) \left(j''+1\right)^2 \left(j''-q\right) \\
 \times \left(q+j''\right) \left(j^2+j-j'' \left(2 q^2+j''-1\right)+j'^2 \left(2 j''-1\right)+j' \left(2 j''-1\right)\right) \\
\end{array}
\right)$ & $ \times$ & $\left[
\begin{array}{ccc}
 j & j''-1 & j' \\
 0 & -q & q \\
 0 & q & -q \\
\end{array}
\right]$ \\ 
 $ + \left(
\begin{array}{c}
 (2 j+1) j' \left(j'+1\right) \left(2 j'+1\right) \left(q-j''\right) \
\left(j''+1\right)^2 \\
 \times \sqrt{-\left(j'^2+j'-q (q+1)\right) \left(q^2+q-\
j''^2+j''\right)} \left(-j^2-j+j'^2+j'+j'' \
\left(2 q+j''-1\right)\right) \\
\end{array}
\right)$ & $ \times$ & $\left[
\begin{array}{ccc}
 j & j''-1 & j' \\
 0 & -q & q \\
 0 & q+1 & -q-1 \\
\end{array}
\right]$ \\ 
 $ + \left(
\begin{array}{c}
 -(2 j+1) \left(j'+1\right)^2 \left(q+j'\right) j'' \left(j''+1\right) \left(2 j''+1\right) \\
 \times \sqrt{\left(-q^2+q+j'^2-j'\right) \
\left(-q^2+q+j''^2+j''\right)} \left(-j^2-j+j'^2+j''^2-(2 q+1) j'+j''\right) \\
\end{array}
\right)$ & $ \times$ & $\left[
\begin{array}{ccc}
 j & j'' & j'-1 \\
 0 & -q & q \\
 0 & q-1 & 1-q \\
\end{array}
\right]$ \\ 
 $ + \left(
\begin{array}{c}
 (2 j+1) q \left(2 j'+1\right) \left(2 j''+1\right)  \sqrt{\left(-q^2+q+j'^2+j'\right) \
\left(-q^2+q+j''^2+j''\right)} \\
 \times \left(j'^4+2 \
j'^3-\left(j^2+j+2 j''^2+2 j''-1\right) \
j'^2-\left(j^2+j+2 j'' \left(j''+1\right)\right) j'+j'' \
\left(j''+1\right) \left(-j^2-j+j''^2+j''\right)\right) \\
\end{array}
\right)$ & $ \times$ & $\left[
\begin{array}{ccc}
 j & j'' & j' \\
 0 & -q & q \\
 0 & q-1 & 1-q \\
\end{array}
\right]$ \\ 
 $ + \left(
\begin{array}{c}
 -(2 j+1) j'^2 \left(-q+j'+1\right) j'' \left(j''+1\right) \left(2 j''+1\right) \\
 \times \sqrt{\left(-q^2+q+j'^2+3 j'+2\right) \
\left(-q^2+q+j''^2+j''\right)} \left(-j^2-j+j'^2+j''^2+2 q+(2 q+3) j'+j''+2\right) \\
\end{array}
\right)$ & $ \times$ & $\left[
\begin{array}{ccc}
 j & j'' & j'+1 \\
 0 & -q & q \\
 0 & q-1 & 1-q \\
\end{array}
\right]$ \\ 
 $ + \left(
\begin{array}{c}
 2 (2 j+1) \left(j'+1\right)^2 \left(j'-q\right) \left(q+j'\right) j'' \
\left(j''+1\right) \\
 \times \left(2 j''+1\right) \left(-j^2-j+j'^2+\left(j\
''\right)^2+j''-j' \left(-2 q^2+2 j''^2+2 j''+1\right)\right) \\
\end{array}
\right)$ & $ \times$ & $\left[
\begin{array}{ccc}
 j & j'' & j'-1 \\
 0 & -q & q \\
 0 & q & -q \\
\end{array}
\right]$ \\ 
 $ + \left(
\begin{array}{c}
 2 (2 j+1) q^2 \left(2 j'+1\right) \left(2 j''+1\right)\\
 \times  \left(j'^4+2 j'^3-\left(j^2+j+2 j''^2+2 j''-1\right) j'^2-\left(j^2+j+2 j'' \left(j''+1\right)\right) j'+j'' \
\left(j''+1\right) \left(-j^2-j+j''^2+j''\right)\right) \\
\end{array}
\right)$ & $ \times$ & $\left[
\begin{array}{ccc}
 j & j'' & j' \\
 0 & -q & q \\
 0 & q & -q \\
\end{array}
\right]$ \\ 
 $ + \left(
\begin{array}{c}
 -2 (2 j+1) j'^2 \left(-q+j'+1\right) \left(q+j'+1\right) j'' \left(j''+1\right) \\
 \times \left(2 j''+1\right) \left(-j^2-j-2 q^2+j'^2+3 \
j''^2+3 j''+j' \left(-2 q^2+2 j''^2+2 j''+3\
\right)+2\right) \\
\end{array}
\right)$ & $ \times$ & $\left[
\begin{array}{ccc}
 j & j'' & j'+1 \\
 0 & -q & q \\
 0 & q & -q \\
\end{array}
\right]$ \\ 
 $ + \left(
\begin{array}{c}
 -(2 j+1) \left(j'+1\right)^2 \left(j'-q\right) j'' \left(j''+1\right) \sqrt{\left(-q+j'-1\right) \left(q+j'\right) \left(j''-q\right) \
\left(q+j''+1\right)} \\
 \times \left(2 j''+1\right) \left(-j^2-j+j'^2+\left(j\
''\right)^2+(2 q-1) j'+j''\right) \\
\end{array}
\right)$ & $ \times$ & $\left[
\begin{array}{ccc}
 j & j'' & j'-1 \\
 0 & -q & q \\
 0 & q+1 & -q-1 \\
\end{array}
\right]$ \\ 
 $ + \left(
\begin{array}{c}
 -(2 j+1) q \left(2 j'+1\right) \sqrt{\left(j'-q\right) \left(q+j'+1\right) \left(j''-q\right) \left(q+j''+1\right)}  \left(2 j''+1\right) \\
 \times \left(j'^4+2 j'^3-\left(j^2+j+2 j''^2+2 j''-1\right) j'^2-\left(j^2+j+2 j'' \left(j''+1\right)\right) j'+j'' \
\left(j''+1\right) \left(-j^2-j+j''^2+j''\right)\right) \\
\end{array}
\right)$ & $ \times$ & $\left[
\begin{array}{ccc}
 j & j'' & j' \\
 0 & -q & q \\
 0 & q+1 & -q-1 \\
\end{array}
\right]$ \\ 
 $ + \left(
\begin{array}{c}
 -(2 j+1) j'^2 \left(q+j'+1\right) j'' \left(j''+1\right) \left(2 j''+1\right) \\
 \times \sqrt{\left(-q^2-q+j'^2+3 j'+2\right) \
\left(j''^2+j''-q (q+1)\right)} \left(-j^2-j+j'^2+j''^2-2 q+(3-2 q) j'+j''+2\right) \\
\end{array}
\right)$ & $ \times$ & $\left[
\begin{array}{ccc}
 j & j'' & j'+1 \\
 0 & -q & q \\
 0 & q+1 & -q-1 \\
\end{array}
\right]$ \\ 
 $ + \left(
\begin{array}{c}
 (2 j+1) j' \left(j'+1\right) \left(2 j'+1\right) \left(q-j''-1\right) j''^2 \\
 \times \sqrt{\left(-q^2+q+j'^2+j'\right) \
\left(-q^2+q+j''^2+3 j''+2\right)} \left(-j^2-j+\left(j'\
\right)^2+j'+\left(j''+1\right) \left(2 q+j''+2\right)\right) \\
\end{array}
\right)$ & $ \times$ & $\left[
\begin{array}{ccc}
 j & j''+1 & j' \\
 0 & -q & q \\
 0 & q-1 & 1-q \\
\end{array}
\right]$ \\ 
 $ + \left(
\begin{array}{c}
 -2 (2 j+1) j' \left(j'+1\right) \left(2 j'+1\right) j''^2 \left(-q+j''+1\right) \\
 \times \left(q+j''+1\right) \left(-j^2-j-2 q^2+j''^2-2 \
q^2 j''+3 j''+j'^2 \left(2 j''+3\right)+j' \left(2 j''+3\
\right)+2\right) \\
\end{array}
\right)$ & $ \times$ & $\left[
\begin{array}{ccc}
 j & j''+1 & j' \\
 0 & -q & q \\
 0 & q & -q \\
\end{array}
\right]$ \\ 
 $ + \left(
\begin{array}{c}
 -(2 j+1) j' \left(j'+1\right) \left(2 j'+1\right) j''^2 \
\left(q+j''+1\right) \\
 \times \sqrt{\left(j'-q\right) \left(q+j'+1\right) \left(-q+j''+1\right) \left(q+j''+2\right)} \left(-j^2-j+j'^2+j'+\left(j''+1\right) \left(-2 q+j''+2\right)\right) \\
\end{array}
\right)$ & $ \times$ & $\left[
\begin{array}{ccc}
 j & j''+1 & j' \\
 0 & -q & q \\
 0 & q+1 & -q-1 \\
\end{array}
\right]$
\end{tabular}}
\ri\}
}
One can check that for $q=0,\,1/2$ we get back the results of the previous subsection.


\section{Asymptotic expansions}\label{sec:Asym}

The goal of this Appendix is to derive an asymptotic formula at large $j$ and $\omega$ for the integrand in the expression \eqref{FqFinalSimp} for $\delta {\cal F}_q$.

\subsection{Small distance expansion of the Green's function}

We start by obtaining a better understanding of the scalar Green's function \eqref{Gq}.  Isolating the phase factor appearing in \eqref{Gq}, we can write
 \es{GqPhaseIsolated}{
  G^q(x, x') = e^{-2 i q \Theta} \widetilde G_q(\tau - \tau', \gamma) \,, \qquad \widetilde G_q(\tau - \tau', \gamma) 
   \equiv \sum_{j = q}^\infty F_{q, j}(\gamma) \frac{e^{- E_{qj} \abs{\tau - \tau'}}}{2 E_{q j}} \,,
 }
where $\gamma$ and $\Theta$ are defined in \eqref{defgamma} and \eqref{defTheta}, respectively, and the energy $E_{qj}$ was defined in \eqref{Eqjfef}.  Note that when $\theta' = \phi' = \tau' = 0$, we have $G_q(x, 0) = \widetilde G_q(\tau, \theta)$.

The Green's function satisfies the differential equation
 \es{DiffEq}{
  \left[\partial_\tau^2 + (\nabla_\mu - i {\cal A}^{q}_\mu)(\nabla^\mu - i {\cal A}^{q,\mu}) 
   - \left( \mu^2  + \frac 14 \right) \right] \widetilde G_q(\tau, \theta) = -\frac{1}{2 \pi \sin \theta} \delta(\tau) \delta(\theta) \,,
 }
where ${\cal A}^{q} =q (1 - \cos \theta) d \phi$.  It is convenient to change variables to 
 \es{VarChange}{
  t = \sinh^2 \frac \tau 2 \,, \qquad s = \sin^2 \frac \theta 2 \,,
 } 
and solve eq.~\eqref{DiffEq} at small $t$ and $s$ (where $t$ and $s$ are considered to be of the same order).  The first few terms in this series expansion are
 \es{GtExpansion}{
  \widetilde G_q(s, t) &= \frac{1}{8 \pi \sqrt{s+t}} + C_q + \frac{\mu_q^2  \sqrt{s + t}}{4 \pi} +
  \left[ A_q t + \frac{s (-2 A_q + C_q (1 + 4 \mu_q^2))}{4}  \right] \\
  &{}+\sqrt{s+t} \frac{(q^2 + 2 \mu_q^2) (2s-t) + 3 \mu_q^4 (s + t)}{36 \pi} \\
  &{}+ B_q t^2 + st \left[ A_q \left( \mu_q^2 - \frac 34 \right) - 3 B_q \right]
    + s^2 \frac{24 B_q - 4 A_q (3 + 4 \mu_q^2) + C_q (9 + 16 q^2  + 40 \mu_q^2 + 16 \mu_q^4)}{64}  \\
  &{}+ \ldots   \,,
 }
where $C_q$, $A_q$, and $B_q$ are integration constants.  At each integer order in the expansion, we have one new integration constant that can be chosen to be the coefficient of $t^n$.  When $s=\gamma = 0$, we have $F_{q, j}(0) = (2 j + 1) / ( 4\pi)$, so 
 \es{Gqt0}{
  \widetilde G_q(0, \tau) &= \frac{1}{8 \pi \sqrt{t}} + C_q + \frac{\mu_q^2  \sqrt{t}}{4 \pi} +
   A_q t  +t^{3/2} \frac{3 \mu_q^4  - 2 \mu_q^2 -  q^2 }{36 \pi} + B_q t^2  + \ldots   \,.
 }
The integration constants can be determined from the spectral decomposition \eqref{GqPhaseIsolated}.  Indeed, for $s = \gamma = 0$, we have 
 \es{SpectralTau}{
  \widetilde G_q(0, \tau) = \frac{1}{4 \pi} \sum_{j = q}^\infty (j+1/2) \frac{e^{- E_{qj} \abs{\tau}}}{E_{q j}} \,,
 }
and this expression can be expanded at small $\tau$ and matched with \eqref{Gqt0}.  The small $\tau$ expansion is subtle, however, because naively expanding the summand at small $\tau$ results in divergent sums.  

In order to only work with absolutely convergent sums, we can first consider the quantity 
 \es{GAnalytic}{
  \widetilde G_q^\text{sub} &= \frac{1}{4 \pi} \sum_{n = q + 1/2}^\infty e^{-n \abs{\tau}} \left[
   1 - \frac{1 + \abs{\tau} n}{2 n^2} (\mu_q^2 - q^2) 
    + \frac{3 + 3 n \abs{\tau} + n^2 \tau^2}{8 n^4} (\mu_q^2 - q^2)^2 + \ldots 
   \right] \\
    &{}+ \frac{1}{4 \pi} \sum_{n=q + 1/2}^\infty \left[\frac{1}{2 n^2} (\mu_q^2 - q^2) + \frac{n^2 \tau^2 - 6}{16 n^4} (\mu_q^2 - q^2)^2 + \ldots \right]  \,.
 }
$\widetilde G_q^\text{sub}$ is a power series in $\mu_q^2 - q^2$. The first line in this expression was obtained by expanding $E_{qj} = \sqrt{(j+1/2)^2 + \mu_q^2 - q^2}$ at small $\mu_q^2 - q^2$ and plugging this expansion into~\eqref{SpectralTau}, thereby obtaining an expansion of $\widetilde G_q(0, \tau)$. The expression in the second line subtracts the first few terms in the small $\tau$ expansion of the first line: at $(\mu_q^2 - q^2)^p$ order it contains terms up to order $\tau^{2(p-1)}$.\footnote{Note that this implies that the $O\le((\mu_q^2 - q^2)^0\ri)$ term is absent in the second line.}  The sums in \eqref{GAnalytic} can be performed analytically, and then expanded at small $t$:
 \es{GAnalyticExpansion}{
  \widetilde G_q^\text{sub} &= \frac{1}{4 \pi} \left[ \frac{1}{2 \sqrt{t}} + \mu_q^2 \sqrt{t} 
   + t^{3/2} \frac{q^2 + 2 \mu_q^2 - 3 \mu_q^4}{9} + \ldots \right] \\
   &{}+\frac{1}{4 \pi} \left[ - q + t \frac{q (1 + 2 q^2 - 6 \mu_q^2)}{6} 
    + \ldots \right]  \,,
 }
where in the first line we isolated the terms non-analytic in $t$, and in the second line we included the terms analytic in $t$.  The non-analytic terms in this expression match exactly the non-analytic terms in \eqref{Gqt0}, so the difference $\widetilde G_q(0, \tau) - \widetilde G_q^\text{sub}$ is analytic in $t$.  In fact, using \eqref{Gqt0} and \eqref{GAnalytic}, it is not hard to see that the difference $\widetilde G_q(0, \tau) - \widetilde G_q^\text{sub}$ and its first few derivatives w.r.t.~$\tau$ (or $t$) are expressed as absolutely convergent sums, so one can expand at small $t$ (or small $\tau$) by expanding the summands.  One finds
 \es{DifferenceExpanded}{
  \widetilde G_q(0, \tau) &- \widetilde G_q^\text{sub} = \frac{1}{4 \pi} \sum_{n=q+1/2}^\infty 
   \Biggl[\left( \frac{n}{E_{\left(n-\frac 12\right)q}} -1\right) 
    + t \left(2 n E_{\left(n-\frac 12\right)q}  - \mu_q^2 + q^2 - 2 n^2 \right)  \\
  &{}+\frac {2t^2}3 \left( n E_{\left(n-\frac 12\right)q}^3 - n E_{\left(n-\frac 12\right)q}
   - n^2 (n^2 - 1) - \frac 12 (3n^2 -1) (\mu_q^2 - q^2) 
    - \frac 38 (\mu_q^2 - q^2)^2 \right) + \ldots  \Biggr]\,.
 }
Adding back \eqref{GAnalyticExpansion}, we can obtain an expression for $\widetilde G_q(0, \tau)$.  It can be checked that this expression can be written more succinctly as
 \es{GreenFinal}{
  \widetilde G_q(0, \tau) &= \frac{1}{4 \pi} \left[ \frac{1}{2 \sqrt{t}} + \mu_q^2 \sqrt{t} 
   + t^{3/2} \frac{q^2 + 2 \mu_q^2 - 3 \mu_q^4}{9} + \ldots \right] \\
   &{}+\frac{1}{4 \pi} \left[ \frac{S_{-1}}{2} + S_1 t + \frac 13(S_3 - S_1) t^2 + \ldots \right]  \,,
 } 
where the first line contains the terms non-analytic in $t$, and in the second line $S_p$ is defined as the zeta-function regularized sum 
 \es{Zeta}{
  S_p \equiv \sum_{j=q}^\infty (2j+1) (E_{qj})^p \,.
 }
In particular,
 \es{SExamples}{
  S_{-1} &= -\frac q2 + \sum_{j= q}^\infty \left[ \frac{2j+1}{E_{qj}} - 2 \right] \,, \\
  S_1 &= \frac{q(1 + 2q^2 - 6 \mu_q^2)}{6} + \sum_{j=q}^\infty \left[(2j+1) E_{qj} - 2 (j+1/2)^2 - \mu_q^2 + q^2 \right] \,, \\
  S_3 &= \frac{q(-7 +10q^2 - 18 q^4 + 30 \mu_q^2 (1 + 2q^2) - 90 \mu_q^4)}{240} \\
   &{}+ \sum_{j=q}^\infty \left[(2j+1) (E_{qj})^3 - 2 (j+1/2)^4 - 3 (\mu_q^2 - q^2) (j+1/2)^2 - \frac 34 (\mu_q^2 - q^2)^2\right] \,,
 }
and so on.  These expressions can easily be evaluated numerically. By comparing \eqref{GreenFinal} to \eqref{Gqt0}, we can extract the constants $C_q$, $B_q$, and $A_q$ appearing in \eqref{GreenFinal}:
 \es{GotABC}{
  C_q = \frac{S_{-1}}{8 \pi} \,, \qquad
   A_q = \frac{S_1}{4 \pi} \,, \qquad B_q = \frac{S_3 - S_1}{12 \pi}  \,.
 }

Note that the constant $C_q$ is the same as that appearing in \eqref{CqDef}.  From \eqref{FInfFinal} and \eqref{solveaq}, it is easy to see that if one tunes to criticality we have 
 \es{AqFinal}{
  C_q = 0\,, \qquad A_q = \frac{{\cal F}_q^\infty}{4 \pi} \,.
 }
The constant $B_q$ can be computed numerically for any given $q$.  We have $B_{1/2} \approx -0.00475009$ and $B_1 \approx -0.01255981$.

With \eqref{GotABC}, we now have a complete expression for the small distance expansion \eqref{GtExpansion} of the Green's function.  This expansion can be developed to higher orders if needed, and in fact the final formulas presented below were obtained after keeping one more order in \eqref{GtExpansion}.

\subsection{UV asymptotic of the scalar kernel}

Let us first examine the scalar kernel $D^q(x, x') = \abs{ G^q(x, x')}^2$.  From \eqref{InvertDSimp} and \eqref{GqPhaseIsolated}, we have
 \es{Dellq}{
  D_j^q(\omega) =2 \pi \int d\tau\, \int d \theta\, \sin \theta \, e^{i \omega \tau} P_{j}(\cos \theta) \widetilde G_q(\tau, \theta)^2   \,.
 }
The large $j$ and $\omega$ behavior of $D_j^q(\omega)$ can be obtained by plugging in the expansion \eqref{GtExpansion} in \eqref{Dellq} and evaluating the integrals {\em provided} that the terms we're Fourier transforming are non-analytic in $s$ and $t$.  (The analytic terms in $s$ and $t$ are {\em not} related in any way to the large $\omega$ and $j$ behavior of $D_j^q(\omega)$.)

So at large $\omega$ and $j$ we have
 \es{DellqExpansion}{
  D_j^q(\omega) &= 2 \pi  \int d\tau\, d \theta\, \sin \theta \, e^{i \omega \tau} P_j(\cos \theta) 
   \Biggl[ \frac{1}{64 \pi^2 (s+t) }
    - \frac{{\cal F}_q^\infty (s-2 t)}{32 \pi^2 \sqrt{s+t}}
      \\
   &{}+  \frac{8 \pi B_q (3 s^2 - 24 st + 8 t^2) + {\cal F}_q^\infty \left[ 32 t^2 \mu_q^2 + 4 st ( - 3 + 8 \mu_q^2) -s^2 (3 + 20 \mu_q^2) \right]}{256 \pi^2 \sqrt{s+t}} + \cdots \Biggr]
 }
In order to evaluate these integrals asymptotically at large $\omega$ and $j$, it is convenient to re-expand the term inside the square brackets in terms of simple functions of 
 \es{XDef}{
  X \equiv \sqrt{2(\cosh \tau - \cos \theta)} = 2 \sqrt{s+t} \,.
 }
For instance, we can write the term inside the square brackets as 
 \es{DReexpansion}{
  D^q_j(\omega) &=  \text{F.T.} \Biggl[ \frac{1}{16 \pi^2 X^2} - \frac{{\cal F}_q^\infty} {192 \pi^2}  \left( \nabla_{S^2}^2 - 2 \partial_\tau^2 \right) X^3 + \frac{B_q}{26880 \pi} \left(8 \partial_\tau^4 - 24 \partial_\tau^2 \nabla_{S^2}^2 + 3 \nabla_{S^2}^4 \right) X^7 \\
  &+ \frac{{\cal F}_q^\infty}{1290240 \pi^2} \left(-38 \partial_\tau^4 + (29 + 96 \mu_q^2) \partial_\tau^2 \nabla_{S^2}^2 - (23  + 24 \mu_q^2) \nabla_{S^2}^4 \right) X^7 + \ldots \Biggr] \,,
 }
 where we introduced the definition of the $S^2 \times \R$ Fourier transform of a function $f(\theta, \tau)$ as
  \es{FTDef}{
   \text{F.T.} \left[ f \right] \equiv  2 \pi  \int d\tau\, d \theta\, \sin \theta \, e^{i \omega \tau} P_j(\cos \theta) f(\theta, \tau) \,.
  }
Eqs.~\eqref{DReexpansion} and \eqref{DellqExpansion} agree up to the order in $s$ and $t$ to which \eqref{DellqExpansion} was valid.

Note that the $S^2 \times \R$ Fourier transform satisfies the properties 
 \es{FTProp}{
  \text{F.T.} \left[ \partial_\tau^2 f \right] &= -\omega^2\, \text{F.T.} \left[ f \right] \,, \\
  \text{F.T.} \left[ \nabla_{S^2}^2 f \right] &= -j(j+1)\, \text{F.T.} \left[ f \right] \,,
 }
which can be derived upon integration by parts twice in \eqref{FTDef}.  These properties, together with the explicit Fourier transforms of powers of $X$ given in Appendix~\ref{FOURIERTRANSFORMS} give
 \es{DellqFinal}{
   D_j^q(\omega) &= D_j^0(\omega) + \frac{{\cal F}_q^\infty}{2 \pi} \frac{(j + \frac 12)^2 - 2 \omega^2}{\left[ (j + \frac 12)^2 + \omega^2 \right]^3} 
    + 6B_q \frac{3 (j + \frac 12)^4 - 24 \omega^2 (j + \frac 12)^2 + 8 \omega^4}{\left[ (j + \frac 12)^2 + \omega^2 \right]^5} \\
   &{}+ \frac{{\cal F}_q^\infty}{8 \pi}  \frac{8(j + \frac 12)^4 (1 - 3 \mu_q^2) + 3 \omega^2 (j + \frac 12)^2 (32 \mu_q^2 - 23) + 25 \omega^4}{\left[ (j + \frac 12)^2 + \omega^2 \right]^5} + O \left( \frac{1}{\left[ (j + \frac 12)^2 + \omega^2 \right]^{7/2} } \right) \,.
 }
Here, $D_j^0(\omega)$ is the quantity defined in \eqref{Dj0Final}.  Asymptotically, at large $\omega$ and $j$,
 \es{DJ0Asymp}{
  D_j^0(\omega) &= \frac{1}{8 \sqrt{(j + \frac 12)^2 + \omega^2}} 
   + \frac{\omega^2 -(j + \frac 12)^2}{64 \left[ (j + \frac 12)^2 + \omega^2 \right]^{5/2}} \\
   &+ \frac{11 (j + \frac 12)^4 - 62 \omega^2 (j + \frac 12)^2 + 11 \omega^4}{1024 \left[ (j + \frac 12)^2 + \omega^2 \right]^{9/2}}
   + O \left( \frac{1}{\left[ (j + \frac 12)^2 + \omega^2 \right]^{7/2} } \right)  \,.
 }

\subsection{The mixed kernel}

Next, we examine the mixed kernel $H^q_\mu(x, x')$ defined in \eqref{KernelsAgain}.  Using $G^q(x, x') = e^{- 2 i q \Theta} \tilde G$, we can write
 \es{DpG}{
  D' G^q &= e^{-2i q \Theta} \left[ d' + i  ({\cal A}^{q})' -2 i q d'\Theta \right] \widetilde G_q \,, \\
  D' G^{q*} &= e^{2i q \Theta} \left[ d' - i ({\cal A}^{q})' + 2 i q d'\Theta \right] \widetilde G_q  \,.
 } 
The definition of $H^q_{\mu}$ then implies
 \es{FqDef}{
  H^q_{\mu}(x, x') = 2 i \left[ {\cal A}^q_\mu (x') -2 \, \partial_\mu' \Theta(x, x') \right] \widetilde G_q(\tau - \tau', \gamma)^2  \,.
 }

From this expression in can be easily seen that $H^q_\tau(x, x') = 0$, which explicitly verifies the argument that $H^{q, \tau}_j(\omega) = H^{q, E}_j(\omega) = 0$ because of $CP$ symmetry.  For $H^{q, B}(\omega)$, we have 
 \es{FqV}{
  H_j^{q, B}(\omega) &=\frac{4 \pi}{2 j + 1} \int d\tau\, d \theta\, d\phi\, \sin \theta\,\lim_{x' \to 0} \left[ e^{i \omega (\tau-\tau')}   \sum_{m=-j}^j Y_{jm}^*(\hat x)   H^q_{\mu'}(x, x')  {\cal Y}^{\mu'}_{jm}(\hat x') \right] \\
   &= -\frac{ 4 \pi i q}{\sqrt{j(j+1)}} \int d\tau\, d \theta\, \sin \theta\, e^{i \omega \tau}   \widetilde G^2_q(\tau, \theta) P_j^1(\cos \theta) \tan \frac{\theta}{2}
 } 

We should use 
 \es{LegPDer}{
  P_j^1(\cos \theta) = - \sin \theta  P_j'(\cos \theta) 
 }
and then 
 \es{FqVAgain}{
  H_j^{q, B}(\omega) 
   &= \frac{ 4 \pi i q}{\sqrt{j(j+1)}} \int d\tau\, d \theta\, \sin \theta\, e^{i \omega \tau}   \widetilde G^2_q(\tau, \theta)  P_j'(\cos \theta) \left( 1 - \cos \theta \right) \,.
 }
We can integrate by parts and obtain  
 \es{FqVAgain2}{
  H_j^{q, B}(\omega) 
   &= -\frac{ 2 \pi i q}{\sqrt{j(j+1)}} \int d\tau\, d \theta\, \sin \theta\, e^{i \omega \tau}  P_j(\cos \theta) \frac{d}{d \cos \theta} \left[2  \widetilde G^2_q(\tau, \theta)  \left( 1 - \cos \theta \right) \right] \\
   &= (-i) \text{F.T.} \left[ \frac{d}{d \cos \theta} \left[2  \widetilde G^2_q(\tau, \theta)  \left( 1 - \cos \theta \right) \right] \right] \,,
 }
where we used the Fourier transform definition in \eqref{FTDef}.  We can use the expansion of the Green's function \eqref{GtExpansion}, re-expanded in terms of $X = \sqrt{2 (\cosh \tau - \cos \theta)}$ as in the scalar kernel case, to write  
 \es{FqVAgain3}{
   H_j^{q, B}(\omega) &= -\frac{i q}{16 \pi^2 \sqrt{j(j+1)}} \text{F.T.} \Biggl[ - \nabla_{S^2} \log X
    + \frac{1}{160} \left(4 \partial_\tau^2 \nabla_{S^2}^2  - \nabla_{S^2}^4 \right) (X^4 \log X) \\
    &{}+\frac{{\cal F}_q^\infty}{180} \left(2 \nabla_{S^2}^4 - 7 \partial_\tau^2 \nabla_{S^2}^2 \right) X^5 + \ldots \Biggr] \,.
 } 
This expression agrees with \eqref{FqVAgain2} in a small $s$ and $t$ expansion up to analytic terms in $s$ and $t$.  Using the Fourier transform properties \eqref{FTProp} and the formulas in Appendix~\ref{FOURIERTRANSFORMS}, we can evaluate the integrals in \eqref{FqVAgain3} and expand at large $\omega$ and $j$.  We obtain
 \es{FApprox}{
  \abs{H_j^{q, B}(\omega)}^2 &= q^2 \frac{(j + \frac 12)^2}{64 \left[ (j + \frac 12)^2 + \omega^2 \right]^3}
   -q^2  \frac{2 (j + \frac 12)^4 - 17 \omega^2 (j + \frac 12)^2 + \omega^4}{256 \left[ (j + \frac 12)^2 + \omega^2 \right]^5} \\
   &{}+ O \left( \frac{1}{\left[ (j + \frac 12)^2 + \omega^2 \right]^{7/2} } \right)   \,.
 }

\subsection{Gauge field kernel}

For the gauge field kernel, we have
 \es{KqSimp}{
  K^q_{\mu\mu'}(x, x') &=2 \partial_\mu   \widetilde G_q(\tau - \tau', \gamma) \partial_{\mu'} \widetilde G_{q} (\tau - \tau', \gamma)
    - 2 \widetilde G_q(\tau - \tau', \gamma) \partial_\mu  \partial_{\mu'}  \widetilde G_{q}(\tau - \tau', \gamma) \\
    &{}- 4  \left[ {\cal A}^q_\mu(x) + 2 \partial_\mu \Theta(x, x')\right] \left[ {\cal A}^q_{\mu'} (x') - 2 \partial_{\mu'} \Theta(x, x')\right]   \widetilde G_q(\tau - \tau', \gamma)^2 \\
    & {}+ 2 g_{\mu \mu'} \delta(x - x') G^q(x, x')   \,. 
 }

It is easiest to start with calculating the Fourier modes of $K^q_{\tau\tau}$ defined by
\es{KtautauDef}{
  K^q_{\tau\tau'} (x, x')  = \int \frac{d\omega}{2 \pi} \sum_{j, m} K_j^{q, \tau\tau}(\omega) Y_{jm}(\hat x) Y_{jm}^*(\hat x') e^{-i \omega(\tau - \tau')}\,.
 }
Using \eqref{KqSimp}, we can extract $K_j^{q, \tau\tau}(\omega)$ from
 \es{Ktauq}{
  K_j^{q, \tau\tau}(\omega) =2 G^q(x, x)+ 2 \pi \int d\tau\, \int d \theta\, \sin \theta \, e^{i \omega \tau} P_{j}(\cos \theta)
   \left[ 2 \widetilde G_q \partial_\tau^2 \widetilde G_q - 2 (\partial_\tau \widetilde G_q)^2  \right]   \,,
 }
where we replaced $\partial_{\tau'} \to -\partial_\tau$ when acting on $\widetilde G_q$.  The first term in \eqref{Ktauq} comes from the last line of \eqref{KqSimp}.   When $q = 0$, we know from \eqref{KFinal} that
 \es{Ktauq0}{
  K_j^{0, \tau\tau}(\omega) = \frac{j(j+1)}{2} D_j^0(\omega)  \,,
 }
where $D_j^0(\omega)$ was given in \eqref{Dj0Final}.

Next, we can calculate the difference $K_j^{q, \tau\tau}(\omega) - K_j^{0, \tau\tau}(\omega)$.  Integrating by parts the first term in \eqref{Ktauq}, we have:
 \es{KtauqAgain}{
  &K_j^{q, \tau\tau}(\omega) - K_j^{0, \tau\tau}(\omega) = 2 \left[ G^q(x, x) - G^0(x, x) \right] \\
   &\qquad{}+ 2 \pi \int d\tau\, \int d \theta\, \sin \theta \, e^{i \omega \tau} P_{j}(\cos \theta)
   \left[- 4  (\partial_\tau \widetilde G_q)^2 - \omega^2 \widetilde G_q^2  + 4  (\partial_\tau \widetilde G_0)^2 + \omega^2 \widetilde G_0^2  \right]   \,.
 }
Since from \eqref{GtExpansion}, $G^q(x, x) - G^0(x, x) = C_q$, we obtain
 \es{KDiff}{
   K_j^{q, \tau\tau}(\omega) -  K_j^{0, \tau\tau}(\omega)
    &= 2 C_q -\omega^2 \left[ D_j^q(\omega) - D_j^0(\omega) \right]  \\
     &-  8 \pi \int d\tau\, \int d \theta\, \sin \theta \, e^{i \omega \tau} P_{j}(\cos \theta)
   \left[  (\partial_\tau \widetilde G_q)^2 -  (\partial_\tau \widetilde G_0)^2  \right]  \,,
 }
where we also used \eqref{Dellq}.

Setting $C_q = 0$, we can write this expression as
 \es{KDiffAgain}{
   K_j^{q, \tau\tau}(\omega) -  K_j^{0, \tau\tau}(\omega)
    &=  -\omega^2 \left[ D_j^q(\omega) - D_j^0(\omega) \right]  -  \text{F.T.} 
   \left[ 4 (\partial_\tau \widetilde G_q)^2 - 4 (\partial_\tau \widetilde G_0)^2  \right]  \,.
 }
Expanding at small $s$ and $t$ and keeping only non-analytic terms, we can write
 \es{KtFT}{
  \text{F.T.} 
   &\left[ 4 (\partial_\tau \widetilde G_q)^2 - 4 (\partial_\tau \widetilde G_0)^2  \right]
    = \frac{1}{8 \pi^2} \text{F.T.} \Biggl[ 
      - \mu_q^2 \nabla_{S^2} \log X
       + {\cal F}_q^\infty \left( \partial_\tau^2 - \nabla_{S^2}^2 \right) X \\
       &{}+ \frac{q^2 + 2 \mu_q^2}{480} \left( 4 \partial_\tau^2 \nabla_{S^2}^2 - \nabla_{S^2}^4 \right) (X^4 \log X) + \frac{\pi B_q}{15} \left(2 \partial_\tau^4 - 7 \partial_\tau^2 \nabla_{S^2}^2 + \nabla_{S^2}^4  \right) X^5 \\
    &{}+\frac{{\cal F}_q^\infty}{120} \left[3 \partial_\tau^4 + 5 (4 \mu_q^2 - 3) \partial_\tau^2 \nabla_{S^2}^2 - 4 \mu_q^2 \nabla_{S^2}^2 \right] X^5 \\
    &{}+ \frac{1}{967680} \biggl[-4 (13 q^2 + 20 \mu_q^2) \partial_\tau^4 \nabla_{S^2}^2
     + (64 q^2 + 92 \mu_q^2) \partial_\tau^2 \nabla_{S^2}^4 \\
     &{}- (10 q^2 + 17 \mu_q^2) \nabla_{S^2}^6\biggr]   (X^8 \log X) 
    + \ldots 
     \Biggr]
 }

Using \eqref{FTProp} and the formulas in Appendix~\ref{FOURIERTRANSFORMS} expanded at large $\omega$ and $j$, we obtain
 \es{KttFinal}{
  K_j^{q, \tau\tau}(\omega) &=K_j^{0, \tau\tau}(\omega)
   + j(j+1) \Biggl[
    \frac{\mu_q^2}{4 \left[(j+1/2)^2 + \omega^2 \right]^{3/2}}
    + \frac{{\cal F}_q^\infty}{2 \pi} \frac{2 (j + \frac 12)^2 - \omega^2}{\left[ (j + \frac 12)^2 + \omega^2 \right]^3} \\
    &{}+ \frac{(\mu_q^2 - 2 q^2) (j + \frac 12)^2 + (11 \mu_q^2 + 8 q^2) \omega^2}
      {32 \left[ (j + \frac 12)^2 + \omega^2 \right]^{7/2}} 
    +   6 B_q \frac{4 (j + \frac 12)^4 - 27 (j + \frac 12)^2 \omega^2 + 4 \omega^4}{\left[ (j + \frac 12)^2 + \omega^2 \right]^{5} } \\
    &{}+\frac{{\cal F}_q^\infty}{8 \pi} 
    \frac{16 (1 -\mu_q^2) (j + \frac 12)^4 + (88 \mu_q^2 - 87) (j + \frac 12)^2 \omega^2 - (1 + 16 \mu_q^2) \omega^4}
      {\left[ (j + \frac 12)^2 + \omega^2 \right]^5} \\
    &{}+ \frac{(68 q^2 + 15\mu_q^2) (j + \frac 12)^4 - 2 (478 q^2 + 111 \mu_q^2) (j + \frac 12)^2 \omega^2
      + (824 q^2 + 519 \mu_q^2) \omega^4}{512 \left[ (j + \frac 12)^2 + \omega^2 \right]^{11/2}}  \\
    &{}+O \left( \frac{1}{\left[ (j + \frac 12)^2 + \omega^2 \right]^4 } \right)  
      \Biggr] \,.
 }

Next, we should consider $K^{q, BB}_j(\omega)$, which can be computed as
 \es{KBBDef}{
  K_j^{q, BB}(\omega) &= 2 G^q(x, x) + \frac{4 \pi}{2 j + 1} \int d\tau\, d \theta\, d\phi\, \sin \theta\,\lim_{x' \to 0} \left[ e^{i \omega (\tau-\tau')}   \sum_{m=-j}^j {\cal Y}^{i}_{jm}(\hat x)   K_{q, ii'}(x, x')  {\cal Y}^{i'*}_{jm}(\hat x') \right] \,, \\
 }
where the first term comes from the contact term in \eqref{KqSimp}.  Plugging things in, we obtain
 \es{GotKBB}{
  K_j^{q, BB}(\omega) &= 2 G^q(x, x) +\frac{4 \pi}{j(j+1)} \int d\tau\, d\theta\, e^{i \omega \tau} \Biggl[
   P_j^1(\cos \theta) \left((\partial_\theta \widetilde G_q)^2 - \widetilde G_q \partial_\theta^2 \widetilde G_q \right) \\
    &\qquad\qquad\qquad\qquad\qquad- \frac{dP_j^1(\cos \theta)}{d \theta} \widetilde G_q \left(\partial_\theta \widetilde G_q 
    + 8 q^2 \csc \theta \sin^4 \frac{\theta}{2} \widetilde G_q \right)
   \Biggr] \,.
 }
We know from \eqref{KFinal} that when $q=0$, 
 \es{KBB0}{
  K^{0, BB}_j(\omega) = \frac{\omega^2 + j^2}{2} D_{j-1}^0(\omega) \,,
 }
so we can calculate $K_j^{q, BB}(\omega) - K_j^{0, BB}(\omega)$.   Using \eqref{LegPDer} and integrating by parts the first term on the second line of \eqref{GotKBB}, we obtain
 \es{KBBDiff}{
  K_j^{q, BB}(\omega) - K_j^{0, BB}(\omega) &= 2C_q + \frac{8 \pi}{j(j+1)} \int d\tau\, d\theta\, e^{i \omega \tau} \Biggl[
   \frac{d P_j(\cos \theta)}{d\theta}  (\partial_\theta \widetilde G_q)^2 \\
     &- \frac{d^2P_j(\cos \theta)}{d \theta^2}  4 q^2 \csc \theta \sin^4 \frac{\theta}{2} \widetilde G_q^2 
      -\frac{d P_j(\cos \theta)}{d\theta}  (\partial_\theta \widetilde G_0)^2
   \Biggr] \,.
 } 
After setting $C_q = 0$ and integrating by parts once the first and third terms under the integral sign and twice the second term, we can write 
 \es{KBBAgain}{
  K_j^{q, BB}(\omega) &= K_j^{0, BB}(\omega) + \frac{1}{8 \pi j(j+1)} \text{F.T.} \Biggl[ 
    -\mu_q^2 \nabla_{S^2}^2 \frac{1}{X^2} + {\cal F}_q^\infty \nabla_{S^2}^2 \frac 1X \\
    &{}+ \frac{1}{24} \left[4 (q^2 + \mu_q^2) \partial_\tau^2 \nabla_{S^2}^2
     + (2 \mu_q^2 - 3 q^2) \nabla_{S^2}^4 \right] (X^2 \log X)\\
    &{}+ \frac{\pi B}{2} \left[4  \partial_\tau^2 \nabla_{S^2}^2
     - \nabla_{S^2}^4 \right] X^3
     + \frac{{\cal F}_q^\infty}{96} \left[(3 - 16 \mu_q^2)  \partial_\tau^2 \nabla_{S^2}^2
     +(8 \mu_q^2 - 3) \nabla_{S^2}^4 \right] X^3 \\
    &{}+ \frac{1}{40320} \biggl[(17 \mu_q^2 - 30 q^2) \nabla_{S^2}^6
     + 2 (8 \mu_q^2 + 69 q^2) \partial_\tau^2 \nabla_{S^2}^4 \\
     &{}- 4 (16 \mu_q^2 + 21 q^2) \partial_\tau^4 \nabla_{S^2}^2  \biggr] (X^6 \log X) 
     +\ldots
  \Biggr]
 }
Using again \eqref{FTProp} and the formulas in Appendix~\ref{FOURIERTRANSFORMS} expanded at large $\omega$ and $j$, we obtain
 \es{KBBFinal}{
  K_j^{q, BB}(\omega) &=K_j^{0, BB}(\omega)
   + \frac{\mu_q^2}{4 \sqrt{(j + \frac 12)^2 + \omega^2}}
   - \frac{{\cal F}_q^\infty}{2 \pi \left[ (j + \frac 12)^2 + \omega^2 \right]} \\
   &{}+ \frac{3 (\mu_q^2 -2 q^2) (j + \frac 12)^2 + (9 \mu_q^2 + 8 q^2) \omega^2}
     {32 \left[ (j + \frac 12)^2 + \omega^2 \right]^{5/2}}  \\
   &{}+ \frac{{\cal F}_q^\infty}{8 \pi} \frac{(8 \mu_q^2 - 3) (j + \frac 12)^2 
      + (3 - 16 \mu_q^2) \omega^2}{\left[ (j + \frac 12)^2 + \omega^2 \right]^3}
   + 6 B_q \frac{(j + \frac 12)^2 - 4 \omega^2}{\left[ (j + \frac 12)^2 + \omega^2 \right]^3} \\
   &{}+ \frac{3 (28 q^2 + \mu_q^2) (j + \frac 12)^4 - 10 (7 \mu_q^2 + 118 q^2) (j + \frac 12)^2 \omega^2
     + (808 q^2 + 459 \mu_q^2) \omega^4}{512 \left[ (j + \frac 12)^2 + \omega^2 \right]^{9/2}}\\
    &{}+ O \left( \frac{1}{\left[ (j + \frac 12)^2 + \omega^2 \right]^3 } \right) \,.
 }

\subsection{Ultraviolet expansion in the $\CP^{N-1}$ model}

The ground state energy in the $\CP^{N-1}$ model can be written as
 \es{Fq}{
  \delta {\cal F}_q = \frac 12 \int \frac{d\omega}{2 \pi} \sum_{j=0}^\infty (2j+1) L^q_j(\omega) \,,
 }
where we defined
 \es{LDef}{
  L^q_j(\omega) \equiv \log \frac{K^{q, \tau\tau}_j(\omega) \left[D^q_j(\omega) 
     K^{q, BB}_j(\omega) + \abs{H^{q, B}_j(\omega)}^2 \right]}
     { D^0_j(\omega) K^{0, \tau\tau}_j(\omega) K^{0, BB}_j(\omega) }  
 }
when $j \geq 1$, and 
 \es{L0Def}{
  L^q_0(\omega) \equiv \log \frac{D^q_0(\omega)}{D^0_0(\omega) } 
 }
when $j=0$.  The large $j$ and $\omega$ behavior of $L^q_j$ can be easily determined from \eqref{DellqFinal}, \eqref{FApprox}, \eqref{KttFinal}, and \eqref{KBBFinal} as well as the asymptotic expansion of $D_j^0(\omega)$ in \eqref{DJ0Asymp}.  We find that for $j>0$ we have
 \es{LAsymp}{
  L^q_j(\omega) &= \frac{8 \mu_q^2}{(j + \frac 12 )^2 + \omega^2 } 
   + \frac{12 {\cal F}_q^\infty}{\pi} \frac{(j + \frac 12 )^2 - \omega^2}{\left[(j + \frac 12 )^2 + \omega^2 \right]^{5/2}} \\
   &{}- \frac{( q^2 + 4 \mu_q^2 (8 \mu_q^2 - 1) ) (j + \frac 12 )^2 
    + 4 (-  q^2+\mu_q^2 (8 \mu_q^2 - 5) ) \omega^2}{2\left[ (j + \frac 12 )^2 + \omega^2 \right]^3} \\
   &{}+144 B_q \frac{3 (j + \frac 12 )^4 - 24 (j + \frac 12 )^2 \omega^2 + 8 \omega^4}{\left[(j + \frac 12 )^2 + \omega^2 \right]^{9/2}} \\
   &{}+\frac{3 {\cal F}_q^\infty}{2 \pi} \frac{(25 - 48 \mu_q^2) (j + \frac 12 )^4 + 3(64 \mu_q^2-55) (j + \frac 12 )^2 \omega^2
    + 20\omega^4 }{\left[ (j + \frac 12 )^2 + \omega^2 \right]^{9/2}} \\
   &{}+ O \left( \frac{1}{\left[ (j + \frac 12)^2 + \omega^2 \right]^3 } \right) \,.
 }
(From the results of this Appendix, one can construct an asymptotic expansion of $L^q_j(\omega)$ that is accurate up to terms that behave as $O(1/\left[(j + \frac 12)^2 + \omega^2 \right]^{7/2})$.  The derivation of the terms not included in \eqref{LAsymp} is straightforward.)  The first term in \eqref{LAsymp} yields a linear UV divergence.  We explain how to regularize this UV divergence in Section~\ref{NUMERICS}.  

When $j=0$, the analog of \eqref{LAsymp} can be obtained from \eqref{DellqFinal} alone:
 \es{L0Asymp}{
  L^q_0(\omega) &= 
    -\frac{8 {\cal F}_q^\infty}{\pi \omega^3} + \frac{32 ({\cal F}_q^\infty + 12 \pi B_q)}{\pi \omega^5}
     - \frac{32 ({\cal F}_q^\infty)^2}{\pi^2 \omega^6} + O(1/\omega^7) \,.
 }

\subsection{Fourier transforms on $S^2 \times \R$}
\label{FOURIERTRANSFORMS}

Here we present some of the Fourier transforms needed in the previous parts of this Appendix.  Recall that the definition of a Fourier transform of a function of $\theta$ and $\tau$ on $S^2 \times \R$ is 
  \es{FTDefAgain}{
   \text{F.T.} \left[ f \right] \equiv  2 \pi  \int d\tau\, d \theta\, \sin \theta \, e^{i \omega \tau} P_j(\cos \theta) f(\theta, \tau) \,.
  }

We can calculate explicitly the Fourier transform of $1/X^{2 \Delta}$, where $X = \sqrt{2 (\cosh\tau - \cos \theta)}$ was defined in \eqref{XDef}.  The calculation proceeds by expanding $1/X^{2 \Delta}$ at large $\tau$ and performing the integrals term by term:
\es{FTX}{
 \text{F.T.} \left[ \frac{1}{X^{2 \Delta}} \right] &=2 \pi \int d\tau d\theta \sin \theta e^{i \omega \tau}
   P_{j}(\cos \theta) \frac{1}{\left[2(\cosh \tau-\cos\theta)\right]^{\Delta}}\\
&= 2 \pi \sum_{n=0}^{\infty}\int d\tau \frac{(\Delta+j)_{n}(\Delta)_{j}(\Delta-1/2)_{n}}{(n+3/2)_{j}(2n+1)!}2^{2n+1}e^{-(\Delta+j+2n)|\tau|+i\omega\tau}\\
&=2 \pi \sum_{n=0}^{\infty}  \frac{(\Delta+j)_{n}(\Delta)_{j}(\Delta-1/2)_{n}}{(n+3/2)_{j}(2n+1)!}2^{2n+2}\frac{\Delta+j+2n}{(\Delta+j+2n)^{2}+\omega^{2}} \,,
}
where $(a)_n$ denotes the Pochhammer symbol.  The sum in the last expression can be performed analytically in terms of hypergeometric functions, but we will not find it helpful to do so here.

When $\Delta = \frac 12 - m$ with $m \geq 0$ an integer, the infinite sum in \eqref{FTX} becomes finite because only the first $m+1$ terms contribute.  We have
 \es{FTXOdd}{
  \text{F.T.} \left[ \frac 1X \right]&= \frac{ 4\pi}{ (j+ \frac 12)^2 + \omega^2} \,, \\
  \text{F.T.} \left[ X \right] &= -\frac{ 8\pi}{\left[ (j+ \frac 32)^2 + \omega^2 \right]\left[ (j- \frac 12)^2 + \omega^2 \right]} \,, \\
  \text{F.T.} \left[ X^3 \right] &= \frac{ 96\pi}{\left[ (j+ \frac 52)^2 + \omega^2 \right]
   \left[ (j+ \frac 12)^2 + \omega^2 \right]\left[ (j- \frac 32)^2 + \omega^2 \right]} \,, 
 }
and so on.  The general formula is
 \es{FTXOddGen}{
  \text{F.T.} \left[ X^{-1+2m} \right] 
   = \frac{ (-1)^m (2m)!  (4 \pi)}{\prod_{k=0}^{m} \left[ (j+ \frac 12 - m + 2k)^2 + \omega^2 \right]} \,.
 }

As can be deduced from Section~\ref{D0KERNEL}, we have
 \es{D0YetAgain}{
  \text{F.T.} \left[ \frac{1}{X^2} \right] = 16 \pi^2 D_j^0(\omega) \,.
 }
By taking derivatives of \eqref{FTX} with respect to $\Delta$ and evaluating the resulting expression at $\Delta = -2m$, where $m \geq 0$ is an integer, we can also calculate 
 \es{FTXEven}{
  \text{F.T.} \left[ \log X \right] &= -\frac{16 \pi^2}{\omega^2 + j^2} D_{j+1}^0(\omega) \,, \\
  \text{F.T.} \left[X^2 \log X \right] &= \frac{16 \pi^2 \times 6}{\left[ \omega^2 + (j-1)^2 \right] \left[ \omega^2 + (j+1)^2 \right]} D_{j+2}^0(\omega) \,, \\
  \text{F.T.} \left[X^4 \log X \right] &= -\frac{16 \pi^2 \times 120}{\left[ \omega^2 + (j-2)^2 \right] \left[ \omega^2 + j^2 \right]\left[ \omega^2 + (j+2)^2 \right]} D_{j+3}^0(\omega) \,, \\
  \text{F.T.} \left[X^6 \log X \right] &= \frac{16 \pi^2 \times 5040}{ \prod_{k=0}^3 \left[ \omega^2 + (j-3 + 2k)^2 \right]} D_{j+4}^0(\omega) \,, \\
   \text{F.T.} \left[X^8 \log X \right] &= -\frac{16 \pi^2 \times 362880}{ \prod_{k=0}^4 \left[ \omega^2 + (j-4 + 2k)^2 \right]} D_{j+5}^0(\omega) \,,
 }
and so on.


\bibliographystyle{ssg}
\bibliography{monopole_scalar}

\begingroup\raggedright\begin{thebibliography}{10}

\bibitem{Polyakov:1975rs}
A.~M. Polyakov, ``{Compact gauge fields and the infrared catastrophe},'' {\em
  Phys.Lett.} {\bf B59} (1975) 82--84.

\bibitem{Wen:1993zza}
X.-G. Wen and Y.-S. Wu, ``{Transitions between the quantum Hall states and
  insulators induced by periodic potentials},'' {\em Phys. Rev. Lett.} {\bf 70}
  (1993) 1501--1504.

\bibitem{Chen:1993cd}
W.~Chen, M.~P. Fisher, and Y.-S. Wu, ``{Mott transition in an anyon gas},''
  {\em Phys. Rev. B} {\bf 48} (1993) 13749--13761,
  \href{http://xxx.lanl.gov/abs/cond-mat/9301037}{{\tt cond-mat/9301037}}.

\bibitem{Sachdev97}
S.~Sachdev, ``{Non-zero temperature transport near fractional quantum Hall
  critical points},'' {\em Phys. Rev. B} {\bf 57} (1998) 7157,
  \href{http://xxx.lanl.gov/abs/cond-mat/9709243}{{\tt cond-mat/9709243}}.

\bibitem{Rantner01}
W.~Rantner and X.-G. Wen, ``{Electron spectral function and algebraic spin
  liquid for the normal state of underdoped high $T_c$ superconductors},'' {\em
  Phys. Rev. Lett.} {\bf 86} (2001) 3871,
  \href{http://xxx.lanl.gov/abs/cond-mat/0010378}{{\tt cond-mat/0010378}}.

\bibitem{Rantner:2002zz}
W.~Rantner and X.-G. Wen, ``{Spin correlations in the algebraic spin liquid:
  Implications for high-$T_c$ superconductors},'' {\em Phys. Rev. B} {\bf 66}
  (2002) 144501, \href{http://xxx.lanl.gov/abs/cond-mat/0201521}{{\tt
  cond-mat/0201521}}.

\bibitem{Motrunich:2003fz}
O.~I. Motrunich and A.~Vishwanath, ``{Emergent photons and new transitions in
  the O(3) sigma model with hedgehog suppression},'' {\em Phys. Rev. B} {\bf
  70} (2004) 075104, \href{http://xxx.lanl.gov/abs/cond-mat/0311222}{{\tt
  cond-mat/0311222}}.

\bibitem{SVBSF}
T.~{Senthil}, A.~{Vishwanath}, L.~{Balents}, S.~{Sachdev}, and M.~P.~A.
  {Fisher}, ``{Deconfined Quantum Critical Points},'' {\em Science} {\bf 303}
  (2004) 1490--1494, \href{http://xxx.lanl.gov/abs/arXiv:cond-mat/0311326}{{\tt
  arXiv:cond-mat/0311326}}.

\bibitem{SBSVF}
T.~{Senthil}, L.~{Balents}, S.~{Sachdev}, A.~{Vishwanath}, and M.~P.~A.
  {Fisher}, ``{Quantum criticality beyond the Landau-Ginzburg-Wilson
  paradigm},'' {\em Phys. Rev. B} {\bf 70} (2004) 144407,
  \href{http://xxx.lanl.gov/abs/arXiv:cond-mat/0312617}{{\tt
  arXiv:cond-mat/0312617}}.

\bibitem{Hermele}
M.~{Hermele}, T.~{Senthil}, M.~P.~A. {Fisher}, P.~A. {Lee}, N.~{Nagaosa}, and
  X.-G. {Wen}, ``{Stability of U(1) spin liquids in two dimensions},'' {\em
  Phys. Rev. B} {\bf 70} (2004) 214437,
  \href{http://xxx.lanl.gov/abs/arXiv:cond-mat/0404751}{{\tt
  arXiv:cond-mat/0404751}}.

\bibitem{Hermele05}
M.~Hermele, T.~Senthil, and M.~P. Fisher, ``{Algebraic spin liquid as the
  mother of many competing orders},'' {\em Phys. Rev. B} {\bf 72} (2005)
  104404, \href{http://xxx.lanl.gov/abs/cond-mat/0502215}{{\tt
  cond-mat/0502215}}.

\bibitem{Ran06}
Y.~{Ran} and X.-G. Wen, ``{Continuous quantum phase transitions beyond Landau's
  paradigm in a large-$N$ spin model},''
  \href{http://xxx.lanl.gov/abs/cond-mat/0609620}{{\tt cond-mat/0609620}}.

\bibitem{Kaul08}
R.~K. Kaul, Y.~B. Kim, S.~Sachdev, and T.~Senthil, ``{Algebraic charge
  liquids},'' {\em Nature Physics} {\bf 4} (2008) 28--31,
  \href{http://xxx.lanl.gov/abs/0706.2187}{{\tt 0706.2187}}.

\bibitem{Kaul:2008xw}
R.~K. Kaul and S.~Sachdev, ``{Quantum criticality of U(1) gauge theories with
  fermionic and bosonic matter in two spatial dimensions},'' {\em Phys. Rev. B}
  {\bf 77} (2008) 155105, \href{http://xxx.lanl.gov/abs/0801.0723}{{\tt
  0801.0723}}.

\bibitem{Sachdev:2010uz}
S.~Sachdev, ``{The landscape of the Hubbard model},'' in {\em {String Theory
  and Its Applications: From meV to the Planck Scale}}, {Proceedings,
  Theoretical Advanced Study Institute in Elementary Particle Physics (TASI
  2010)}, pp.~559--620, {World Scientific}, 2010.
\newblock \href{http://xxx.lanl.gov/abs/1012.0299}{{\tt 1012.0299}}.

\bibitem{Appelquist:1981vg}
T.~Appelquist and R.~D. Pisarski, ``{High-Temperature Yang-Mills Theories and
  Three-Dimensional Quantum Chromodynamics},'' {\em Phys. Rev. D} {\bf 23}
  (1981) 2305.

\bibitem{Appelquist:1986fd}
T.~W. Appelquist, M.~J. Bowick, D.~Karabali, and L.~Wijewardhana,
  ``{Spontaneous Chiral Symmetry Breaking in Three-Dimensional QED},'' {\em
  Phys. Rev. D} {\bf 33} (1986) 3704.

\bibitem{Appelquist:1988sr}
T.~Appelquist, D.~Nash, and L.~Wijewardhana, ``{Critical behavior in
  $(2+1)$-dimensional QED},'' {\em Phys. Rev. Lett.} {\bf 60} (1988) 2575.

\bibitem{Appelquist:1981sf}
T.~Appelquist and U.~W. Heinz, ``{Three-dimensional $O(N)$ theories at large
  distances},'' {\em Phys. Rev. D} {\bf 24} (1981) 2169.

\bibitem{coleman1988aspects}
S.~Coleman, {\em Aspects of symmetry: selected Erice lectures}.
\newblock Cambridge University Press, 1988.

\bibitem{Read:1990zza}
N.~Read and S.~Sachdev, ``{Spin-Peierls, valence-bond solid, and Neel ground
  states of low-dimensional quantum antiferromagnets},'' {\em Phys. Rev. B}
  {\bf 42} (1990) 4568--4589.

\bibitem{Read:1989zz}
N.~Read and S.~Sachdev, ``{Valence-bond and spin-Peierls ground states of
  low-dimensional quantum antiferromagnets},'' {\em Phys. Rev. Lett.} {\bf 62}
  (1989) 1694--1697.

\bibitem{Murthy:1989ps}
G.~Murthy and S.~Sachdev, ``{Action of hedgehog instantons in the disordered
  phase of the $(2+1)$-dimensional $\CP^{N-1}$ model},'' {\em Nucl.Phys.} {\bf
  B344} (1990) 557--595.

\bibitem{2009PhRvB..80r0414L}
J.~{Lou}, A.~W. {Sandvik}, and N.~{Kawashima}, ``{Antiferromagnetic to
  valence-bond-solid transitions in two-dimensional SU($N$) Heisenberg models
  with multispin interactions},'' {\em Phys. Rev. B} {\bf 80} (2009) 180414,
  \href{http://xxx.lanl.gov/abs/0908.0740}{{\tt 0908.0740}}.

\bibitem{2012PhRvL.108m7201K}
R.~K. {Kaul} and A.~W. {Sandvik}, ``{Lattice Model for the SU($N$) N{\'e}el to
  Valence-Bond Solid Quantum Phase Transition at Large $N$},'' {\em Phys. Rev.
  Lett.} {\bf 108} (2012) 137201, \href{http://xxx.lanl.gov/abs/1110.4130}{{\tt
  1110.4130}}.

\bibitem{2013PhRvL.111h7203P}
S.~{Pujari}, K.~{Damle}, and F.~{Alet}, ``{N{\'e}el-State to Valence-Bond-Solid
  Transition on the Honeycomb Lattice: Evidence for Deconfined Criticality},''
  {\em Phys. Rev. Lett.} {\bf 111} (2013) 087203,
  \href{http://xxx.lanl.gov/abs/1302.1408}{{\tt 1302.1408}}.

\bibitem{2013PhRvL.111m7202B}
M.~S. {Block}, R.~G. {Melko}, and R.~K. {Kaul}, ``{Fate of CP$^{N-1}$ Fixed
  Points with q Monopoles},'' {\em Phys. Rev. Lett.} {\bf 111} (2013) 137202,
  \href{http://xxx.lanl.gov/abs/1307.0519}{{\tt 1307.0519}}.

\bibitem{2015arXiv150205128K}
R.~K. {Kaul} and M.~S. {Block}, ``{Numerical studies of various N{\'e}el-VBS
  transitions in SU(N) anti-ferromagnets},'' {\em J. Phys. Conf. Series} {\bf
  640} (2015) 012041, \href{http://xxx.lanl.gov/abs/1502.05128}{{\tt
  1502.05128}}.

\bibitem{Borokhov:2002ib}
V.~Borokhov, A.~Kapustin, and X.-k. Wu, ``{Topological disorder operators in
  three-dimensional conformal field theory},'' {\em JHEP} {\bf 0211} (2002)
  049, \href{http://xxx.lanl.gov/abs/hep-th/0206054}{{\tt hep-th/0206054}}.

\bibitem{Borokhov:2002cg}
V.~Borokhov, A.~Kapustin, and X.-k. Wu, ``{Monopole operators and mirror
  symmetry in three dimensions},'' {\em JHEP} {\bf 0212} (2002) 044,
  \href{http://xxx.lanl.gov/abs/hep-th/0207074}{{\tt hep-th/0207074}}.

\bibitem{Metlitski:2008dw}
M.~A. Metlitski, M.~Hermele, T.~Senthil, and M.~P. Fisher, ``{Monopoles in
  $\CP^{N-1}$ model via the state-operator correspondence},'' {\em Phys. Rev.
  B} {\bf 78} (2008) 214418, \href{http://xxx.lanl.gov/abs/0809.2816}{{\tt
  0809.2816}}.

\bibitem{Pufu:2013vpa}
S.~S. Pufu, ``{Anomalous dimensions of monopole operators in three-dimensional
  quantum electrodynamics},'' {\em Phys. Rev. D} {\bf 89} (2014) 065016,
  \href{http://xxx.lanl.gov/abs/1303.6125}{{\tt 1303.6125}}.

\bibitem{Dyer:2013fja}
E.~Dyer, M.~Mezei, and S.~S. Pufu, ``{Monopole Taxonomy in Three-Dimensional
  Conformal Field Theories},'' \href{http://xxx.lanl.gov/abs/1309.1160}{{\tt
  1309.1160}}.

\bibitem{ScalarQED}
E.~Dyer, M.~Mezei, and S.~Pufu, unpublished.

\bibitem{Pufu:2013eda}
S.~S. Pufu and S.~Sachdev, ``{Monopoles in $2 + 1$-dimensional conformal field
  theories with global U(1) symmetry},'' {\em JHEP} {\bf 1309} (2013) 127,
  \href{http://xxx.lanl.gov/abs/1303.3006}{{\tt 1303.3006}}.

\bibitem{Sachdev:2012tj}
S.~Sachdev, ``{Compressible quantum phases from conformal field theories in 2+1
  dimensions},'' {\em Phys. Rev. D} {\bf 86} (2012) 126003,
  \href{http://xxx.lanl.gov/abs/1209.1637}{{\tt 1209.1637}}.

\bibitem{Iqbal:2014cga}
N.~{Iqbal}, ``{Monopole correlations in holographically flavored liquids},''
  {\em Phys. Rev. D} {\bf 91} (2015) 106001,
  \href{http://xxx.lanl.gov/abs/1409.5467}{{\tt 1409.5467}}.

\bibitem{Borokhov:2003yu}
V.~Borokhov, ``{Monopole operators in three-dimensional ${\cal N}=4$ SYM and
  mirror symmetry},'' {\em JHEP} {\bf 0403} (2004) 008,
  \href{http://xxx.lanl.gov/abs/hep-th/0310254}{{\tt hep-th/0310254}}.

\bibitem{Benna:2009xd}
M.~K. Benna, I.~R. Klebanov, and T.~Klose, ``{Charges of Monopole Operators in
  Chern-Simons Yang-Mills Theory},'' {\em JHEP} {\bf 1001} (2010) 110,
  \href{http://xxx.lanl.gov/abs/0906.3008}{{\tt 0906.3008}}.

\bibitem{Gustavsson:2009pm}
A.~Gustavsson and S.-J. Rey, ``{Enhanced ${\cal N}=8$ Supersymmetry of ABJM
  Theory on $\R^8$ and $\R^8/\Z_2$},''
  \href{http://xxx.lanl.gov/abs/0906.3568}{{\tt 0906.3568}}.

\bibitem{Benini:2009qs}
F.~Benini, C.~Closset, and S.~Cremonesi, ``{Chiral flavors and M2-branes at
  toric CY4 singularities},'' {\em JHEP} {\bf 1002} (2010) 036,
  \href{http://xxx.lanl.gov/abs/0911.4127}{{\tt 0911.4127}}.

\bibitem{Benini:2011cma}
F.~Benini, C.~Closset, and S.~Cremonesi, ``{Quantum moduli space of
  Chern-Simons quivers, wrapped D6-branes and AdS4/CFT3},'' {\em JHEP} {\bf
  1109} (2011) 005, \href{http://xxx.lanl.gov/abs/1105.2299}{{\tt 1105.2299}}.

\bibitem{Aharony:2015pla}
O.~Aharony, P.~Narayan, and T.~Sharma, ``{On monopole operators in
  supersymmetric Chern-Simons-matter theories},'' {\em JHEP} {\bf 05} (2015)
  117, \href{http://xxx.lanl.gov/abs/1502.00945}{{\tt 1502.00945}}.

\bibitem{Wu:1976ge}
T.~T. Wu and C.~N. Yang, ``{Dirac Monopole Without Strings: Monopole
  Harmonics},'' {\em Nucl.Phys.} {\bf B107} (1976) 365.

\bibitem{Wu:1977qk}
T.~T. Wu and C.~N. Yang, ``{Some properties of monopole harmonics},'' {\em
  Phys. Rev. D} {\bf 16} (1977) 1018--1021.

\bibitem{AgmonThesis}
N.~B. Agmon, ``{In Search of Spherically Non-Symmetric Saddle Points of the
  $\CP^{N-1}$ Model in $2+1$ Dimensions}.'' Junior Paper, Princeton University,
  2016.

\bibitem{zinn2002quantum}
J.~Zinn-Justin, {\em Quantum Field Theory and Critical Phenomena}.
\newblock International series of monographs on physics. Clarendon Press, 2002.

\bibitem{Rattazzi:2008pe}
R.~Rattazzi, V.~S. Rychkov, E.~Tonni, and A.~Vichi, ``{Bounding scalar operator
  dimensions in 4D CFT},'' {\em JHEP} {\bf 0812} (2008) 031,
  \href{http://xxx.lanl.gov/abs/0807.0004}{{\tt 0807.0004}}.

\bibitem{Simmons-Duffin:2015qma}
D.~Simmons-Duffin, ``{A Semidefinite Program Solver for the Conformal
  Bootstrap},'' {\em JHEP} {\bf 06} (2015) 174,
  \href{http://xxx.lanl.gov/abs/1502.02033}{{\tt 1502.02033}}.

\end{thebibliography}\endgroup

\end{document}